\documentclass[onecolumn,preprint,prd,tightenlines]{revtex4}
\usepackage{amsmath}

\newcommand{\CA}{{\cal A}}
\begin{document}

\title{$SU(3)$ Decay Amplitudes of Pentaquarks into Decuplet Baryons}

\author{Benjam\'\i{}n Grinstein and Mikhail A. Savrov}
\affiliation{Department of Physics, University of California at San Diego,
  La Jolla, CA 92093\vspace{4pt} }
\date{\today}
\preprint{UCSD/PTH 04-16}

\begin{abstract}
Evidence for the existence of exotic, pentaquark baryons was first
found in the invariant mass spectrum of a baryon-meson pair, with the
baryon in the octet of flavor $SU(3)$. Exotic pentaquarks may also
decay into members of the spin-$\frac32$ baryon decuplet. We
calculate relations between decay rates of the exotic pentaquark into
a baryon decuplet and an octet of pseudoscalar mesons based on
$SU(3)$ flavor symmetry, including the leading flavor symmetry
breaking term. We consider all possible representations for an exotic
pentaquark, namely, that it belongs to either a $\mathbf{\overline {10}}$,
$\mathbf{27}$ or $\mathbf{35}$ representation of flavor $SU(3)$.  In
addition, we use the approximate $SU(6)$ spin-flavor symmetry of
baryons to derive relations between the reduced matrix elements of our
calculation and those of an existing analogous calculation for decays
into the baryon octet. By comparing several decay rates of exotics
into both octet and decuplet baryons, our results could be used to
elucidate the $SU(3)$ nature of pentaquarks.
\end{abstract}

\maketitle 

\section{Introduction}\label{intro}
Recent evidence for pentaquark states~(see, e.g.~\cite{theta}) has
generated excitement and a resurgence of interest in hadron
spectroscopy. For the recent review of the experimental status of
pentaquark searches see~\cite{hicks} and references therein. The very
existence of these exotic states provides a challenge for
theorists. It is fair to say that our understanding of the mechanism
by which strong forces bind quarks into baryons is crude. There is no
theoretical consensus on the nature of the pentaquark
states or even as to whether they are predicted by QCD\cite{theories},
let alone on the production mechanisms\cite{production} of exotics in
different experimental setups.

Flavor symmetry may help us gain some insights into the nature of the
forces that bind quarks. Pentaquarks form one or more multiplet of
states that belong to an irreducible representation of $SU(3)$-flavor
symmetry. For a state with the quantum numbers of four quarks and one
antiquark the possible exotic representations are $\mathbf{\overline {10}}$,
$\mathbf{27}$ or $\mathbf{35}$. Which of these representations
is found in nature is determined by strong interaction
dynamics. Therefore, a determination of which $SU(3)$ multiplet
observed pentaquarks fall into would advance, if only minimally, our
understanding of the forces that bind them. With only a few members of
a mutiplet observed, distinguishing between representations is
challenging. $SU(3)$ relations indicate different patterns of decays
for different representations.

The implicit assumption for an $SU(3)$-symmetry analysis is the
approximate validity of the symmetry. Symmetry breaking effects
originate from difference in light quark masses (and the fact that the
largest, $m_s$ is not negligibly small relative to the hadronic
scale). Electromagnetic interactions also break the symmetry, but the
effect is usually much smaller. Our analysis incorporates symmetry
breaking effects to first order in $m_s$. A wealth of evidence
accumulated over 40 years shows that $SU(3)$ flavor symmetry relations
hold to about 30\% accuracy. Therefore, our relations for pentaquark
decays are expected to be valid to a precision of approximately
$(0.3)^2\sim$~10\%. Of course, the analysis is a prerequisite for a
test of the validity of $SU(3)$ symmetry in this new hadronic realm. 

Baryons exhibit yet another approximate symmetry\cite{bib:su6}:
$SU(6)_c$ (the "c" stands for "contracted"). The origin of the
symmetry is not completely understood. With some technical assumptions
it has been shown that $SU(6)_c$ appears as a symmetry of baryons in
the large $N_c$ expansion of QCD\cite{dashen}.  Implications of the
symmetry, including $SU(6)_c$ breaking effects, have been investigated
in detail.  The $1/N_c$ expansion enables one to investigate
systematically these symmetry breaking effects. The comparison with
experiment shows that this expansion works fairly
well~\cite{largeN}. Hence it is reasonable to expect that $SU(6)_c$
holds as an approximate symmetry for properties of pentaquarks.

In this paper we calculate decay amplitudes of pentaquarks into
decuplet baryons and pseudoscalar mesons in terms of several reduced
matrix elements using $SU(3)$-flavor symmetry including first order
symmetry breaking. Together with the amplitudes for the decays into
octet baryons in Ref.~\cite{paper} they give the complete picture of
decay amplitudes into ground state baryons. Previous work had
considered the symmetry relations that follow in the $SU(3)$ symmetry
limit\cite{Huang:2003we} ignoring symmetry breaking effects.
Furthermore we use $SU(6)_c$ symmetry to give all of the reduced
matrix elements for all of the exotic multiplets decaying into both
octet and decuplet baryons in terms of four independent
parameters. To this end we assume that the exotic pentaquark
multiplets are all members of the $700$ irreducible representation of
$SU(6)$. We have not investigated relations that arise from assuming
the exotics fall into other irreducible representations of $SU(6)$. We
have only included flavor symmetry breaking of $SU(6)$, but neglected
symmetry breaking suppressed by $1/N_c$ that includes, for example,
spin-symmetry breaking. That is, we work consistently to leading order
in $1/N_c$ and to first order in $m_s$. 

The paper is organized as follows.  The results of the calculation are
presented in Secs.~\ref{10table},~\ref{27table}, and~\ref{35table},
and the brief Sec.~\ref{howto} explains the contents of those tables.
The method used to perform the calculation leading to those tables is
described in section~\ref{su3}.  The $SU(6)$ calculation relating the
reduced matrix elements is presented in section~\ref{su6}. The reader
who needs to use the results may safely skip Secs.~\ref{su3}
and~\ref{su6}, and can first refer to Sec.~\ref{howto} for
instructions on how to use the tables, and then consult them in
Secs.~\ref{10table}--\ref{35table}.

\section{How to read our results}\label{howto}
We here briefly explain how to use the tables in
Secs.~\ref{10table},~\ref{27table}, and~\ref{35table}, how to further
sharpen those results using the $SU(6)$ symmetry relations in
Sec.~\ref{su6} and give some examples.

Consider the pentaquark decay $ \Xi^{--}_{\overline{10}}\to \Xi^{*-}\pi^-$, with
exotic final state flavor $ssdd\bar u$. From the table in
Sec.~\ref{10table} we read that the amplitude for this process is
$\CA( \Xi^{--}_{\overline{10}}\to
\Xi^{*-}\pi^-)=-5\sqrt{3}{{\gamma^*_{\overline{10}}}}$. 
Similarly, $\CA(
\Xi^{0}_{\overline{10}}\to \Xi^{*-}\pi^+)=5{{\gamma^*_{\overline{10}}}}$. It
follows immediately that
\begin{equation}
\label{eq:sample1}
\Gamma(\Xi^{--}_{\overline{10}}\to \Xi^{*-}\pi^-) =3
\Gamma(\Xi^{0}_{\overline{10}}\to \Xi^{*-}\pi^+).
\end{equation}

This relation is a consequence of isospin symmetry.  Relations that do
not follow solely from isospin must account for differences in
kinematics. These arise from two sources, phase space and momentum
dependence of the reduced matrix elements in the amplitudes. In
addition, since these differences are a consequence of $SU(3)$
breaking, one must also account for $SU(3)$ breaking in the amplitude,
and our results include symmetry breaking to linear order in the
symmetry breaking parameter, $m_s$.  

While accounting for phase space differences is trivial, for the
kinematic dependence of the amplitudes we need further understanding
of the dynamics. Soft pion theorems require that pseudoscalar mesons
are derivatively coupled. Hence the amplitude is at least linear in
the momentum, and we make the assumption that this linear coupling is
present. In particular, this means that our results are a simultaneous
expansion in the $SU(3)$ symmetry breaking parameter and in the
momentum of the pseudoscalar meson. We see then that rates are
proportional to the third power of momentum. The analysis in
section~\ref{su6} makes this explicitly clear since it introduces the
coupling of the pseudoscalar mesons with an axial current as an
interpolating field. For example, from the same table we have also
$\CA(\Xi^{--}_{\overline{10}}\to
\Sigma^{*-}K^-)=5\sqrt{3}{{\gamma^*_{\overline{10}}}}$ and hence
\begin{equation}
\label{eq:sample2}
\frac{1}{|\vec p|^3}\Gamma(\Xi^{--}_{\overline{10}}\to \Sigma^{*-}K^-)
=\frac{1}{|\vec p|^3}\Gamma(\Xi^{--}_{\overline{10}}\to \Xi^{*-}\pi^-) 
=\frac{3}{|\vec p|^3} \Gamma(\Xi^{0}_{\overline{10}}\to \Xi^{*-}\pi^+),
\end{equation}
where it is understood that $\vec p$ stands for the momentum of the
final state particle in the relevant reaction.  Some decays,
particularly when the final state includes a $K$-meson, are not
kinematically allowed and should not be included in the relations. In
this example $m_K+m_{\Sigma^{*}}=1880$~MeV, and a candidate for
$\Xi^{--}_{\overline{10}}$ has been reported\cite{xi} with mass
$1862$~MeV, so the first relation in Eq.~(\ref{eq:sample2}) above
becomes meaningless.

We include in the tables all possible decay modes consistent with
flavor conservation, regardless of whether the process is
kinematically allowed. The amplitudes for kinematically forbidden
modes may be useful in computing off-shell amplitudes.

Referring now to section~\ref{su6} we see, from Eq.~(\ref{su6results}),
that $SU(6)_c$ implies ${{\gamma^*_{\overline{10}}}}=0$. So, to this order the
decay rates above vanish. This means, for example, that compared to
the decay rate for $\Xi^{--}_{\overline{10}}$ into members of the baryon
octet, which are allowed in the symmetry limit, the decays into
members of the decuplet are suppressed by at least two powers of $m_s$:
\begin{equation}
\label{chimmresult}
\frac{\frac{1}{|\vec p|^3}\Gamma(\Xi^{--}_{\overline{10}}\to
  \Xi^{*-}\pi^-)}{\frac{1}{|\vec p|^3}\Gamma(\Xi^{--}_{\overline{10}}\to
  \Xi^{-}\pi^-)}\lesssim(0.3)^2\sim0.1.
\end{equation}
This result is peculiar to the $\mathbf{\overline{10}}$ multiplet. In
fact the roles of allowed and suppressed decays are inverted in
Eq.~(\ref{chimmresult}) if the $\Xi^{--}$ belongs to the $\mathbf{35}$
multiplet, and the ratio is order of unity if in the $\mathbf{27}$. We
emphasize that we only retain terms of leading order in $1/N_c$. In
particular, terms that break $SU(6)$ through spin operators are
neglected. 

In the examples of Eq.~(\ref{eq:sample2}), the amplitudes were all
given in terms of a single reduced matrix element. This is not
generally the case. Consider, for example, decays of a $\Xi^{'--}$ that
belongs to the $\mathbf{27}$:
\begin{align}
\frac{1}{|\vec p|^3}\Gamma(\Xi^{'--}_{27}\to \Sigma^{*-}K^-)
&=|-\sqrt{2/3}\alpha^*_{27}+4\sqrt{2/3} \delta^*_{27}
        -2\sqrt{2/3} \epsilon^*_{27} -24\sqrt{6}\zeta^*_{27}|^2\\
\frac{1}{|\vec p|^3}\Gamma(\Xi^{'--}_{27}\to \Xi^{*-}\pi^-) 
&=|-\sqrt{2/3}\alpha^*_{27}+4\sqrt{2/3} \delta^*_{27}
        -2\sqrt{2/3} \epsilon^*_{27} +24\sqrt{6}\zeta^*_{27}|^2.
\end{align}
We see that for the $\mathbf{27}$ the amplitudes differ at order $m_s$. 

There is an implicit spin-wavefunction factor in our amplitudes. We
have omitted this from our expressions, since it is straightforward to
re-introduce it. For example, the amplitude for $
\Xi^{--}_{\overline{10}}\to \Xi^{*-}\pi^-$ discussed at the opening of this
section involves a spin-$\frac12$ to a spin-$\frac12$ plus spin-0
decay. The amplitude has a factor of 
\begin{equation}
\frac1{\sqrt2}\chi( \Xi^{*-})^\dagger[\hat{\vec p}\cdot \vec \sigma]
\chi(\Xi^{--}_{\overline{10}})
= \frac1{\sqrt2}\left(\chi( \Xi^{*-})^\dagger_\uparrow
\chi(\Xi^{--}_{\overline{10}})^\uparrow-\chi(
\Xi^{*-})^\dagger_\downarrow
\chi(\Xi^{--}_{\overline{10}})^\downarrow\right),
\label{eq:spin1}
\end{equation}
where $\chi$ is a bi-spinor, $\vec \sigma $ are Pauli matrices,
$\hat{\vec p}$ is a unit vector in the direction of the pion momentum,
and the last equality follows from assuming $\vec p$ is along the
$z$-axis. In fact, with $\vec p$ along the $z$-axis, the same factor
occurs for transitions between spin-$\frac12$ and spin-$\frac32$
baryons, with the understanding that only the middle two components of
the four component spin-$\frac32$ spinor contribute.  The case of
spin-$\frac32\to$~spin-$\frac32$ is slightly different. The factor in
these amplitudes is
\begin{equation}
\frac1{\sqrt{20}}\left(
3\chi^\dagger_{\frac32}\chi'_{\frac32}
+\chi^\dagger_{\frac12}\chi'_{\frac12}
-\chi^\dagger_{-\frac12}\chi'_{-\frac12}
-3\chi^\dagger_{-\frac32}\chi'_{-\frac32}
\right),
\label{eq:spin2}
\end{equation}
where, again, we have specialized to the case of $\vec p$ along the
$z$-axis, and $\chi$ and $\chi'$ are four component spinors.

\section{$SU(3)$ Calculation}\label{su3}

In this section the calculation of $SU(3)$ relations is described. The
computational algorithm is the same as that implemented
in~\cite{paper}, where we refer the interested reader for details.

\subsection{$({\bf{\overline{10}}},{\frac12})$ Decay}\label{sec10b}
 
The tensor decomposition 
\begin{equation}
\label{eq:decayprod}
{\bf{8}}\otimes {\bf{10}}={\bf{8}} \oplus {\bf{10}} \oplus {\bf{27}} \oplus {\bf{35}}
\end{equation}
shows that the decay of a $\overline{{\bf{10}}}$ pentaquark into baryon
decuplet and octet of pseudoscalar mesons is forbidden in the limit of
$SU(3)$ flavor symmetry. The first contribution to the decay amplitude
is due to the flavor breaking piece in the strong interactions 
Hamiltonian. So, the matrix element to be calculated is
\begin{equation}
\langle \mathbf{8} \otimes \mathbf{10} | O^i_j(T_8)^j_i| \overline{\mathbf{10}} \rangle . \label{eq:matelement1}
\end{equation}
Here $O^i_j$ is a decay operator due to the chiral symmetry breaking
term in QCD Hamiltonian that transforms like the $T_8$ component of an
octet tensor under flavor $SU(3)$. Vector $\langle \mathbf{8} \otimes
\mathbf{10} |$ is the direct product of the pseudoscalar meson octet and
baryon decuplet states. Vector $| \overline{\mathbf{10}} \rangle$
stands for a $\overline{\mathbf{10}}$ pentaquark. Taking the product
of the  tensor
decomposition Eq.~(\ref{eq:decayprod}) with its conjugate,
\begin{equation}
\label{eq:8x10b}
{\bf{8}}  \otimes \overline{{\bf{10}}} = {\bf{8}} \oplus \overline{{\bf{10}}} \oplus {\bf{27}} \oplus \overline{{\bf{35}}} 
\end{equation}
we see that there are only two irreducible matrix elements
parameterizing Eq.~(\ref{eq:matelement1}), corresponding to the
product of two {\bf{8}}'s and of two {\bf{27}}'s in the
meson$\times$baryon and $T_8\times$pentaquark tensor
decompositions. The explicit form of the meson ($M^i_j$), baryon ($B^i_j$),
and $\overline{{\bf{10}}}$~pentaquark ($P^i_j$)  tensors is given by Eqs.~(5)
and~(6) in~\cite{paper}. (Note that in the present paper we use symbols
$P$ for pentaquark states and $D$ for decuplet states.) The independent
components of the completely symmetric baryon decuplet tensor are
\begin{eqnarray}
&&
D^{111} = \Delta^{++}, \qquad
D^{112} = \frac{\Delta^+}{\sqrt{3}}, \qquad
D^{122} = \frac{\Delta^0}{\sqrt{3}}, \qquad
D^{222} = \Delta^-, 
\nonumber \\  &&
D^{113} = \frac{\Sigma^{*+}}{\sqrt{3}}, \qquad
D^{123} = \frac{\Sigma^{*0}}{\sqrt{6}} , \qquad
D^{223} = \frac{\Sigma^{*-}}{\sqrt{3}}, 
\nonumber \\  &&
D^{133} = \frac{\Xi^{*0}}{\sqrt{3}}, \qquad
D^{113} = \frac{\Xi^{*-}}{\sqrt{3}}, 
\nonumber \\  &&
D^{333} = \Omega^{-}.
\end{eqnarray}
The explicit form of the irreducible ${\bf{8}}$ and ${\bf{27}}$
representations in the decomposition of the direct product of meson and
baryon tensors is
\begin{align}
 8^i_j&=\epsilon^{irs}(M^{\dagger})^q_r(D^{\dagger})_{qsj} \nonumber\\
 27^{ik}_{jl}&=\epsilon^{rs(i}(M^\dagger)^{k)}_r(D^\dagger)_{sjl}-\frac{1}{5}\delta^{(i}_{(j}8^{k)}_{l)}.
\label{eq:irrep1}
\end{align}
Here parentheses stand for symmetrization. The {\bf{27}} tensor is
symmetric both in upper $(ik)$ and lower $(jl)$ indices.  Hermitian
conjugates are used because both tensors are wavefunctions of $bra$
vectors and therefore transform as complex conjugate under $SU(3)$
flavor group. Analogously, the corresponding {\bf{8}} and {\bf{27}} in
the decomposition of the product of $T_8$ and pentaquark tensor are
\begin{align}
 8^i_j&=\frac{1}{\sqrt{2}}\epsilon^{irs}(T_8)^q_r P_{jsq} \nonumber \\
 27^{ik}_{jl}&=\epsilon^{rs(i}(T_8)^{k)}_rP_{sjl}-\frac{1}{5}\delta^{(i}_{(j}8^{k)}_{l)}.
\label{eq:irrep2}
\end{align}
The extra factor of $1/\sqrt{2}$ is a convenient, although largely
irrelevant, normalization. Finally, we write the matrix
element in Eq.~(\ref{eq:matelement1}) in terms of two parameters
(reduced matrix elements): 
\begin{equation}
\langle \mathbf{8} \otimes \mathbf{10} | O^i_j(T_8)^j_i |
\overline{\mathbf{10}}
\rangle={{}\beta^*_{\overline{10}}}\cdot(8^{\dagger})^i_j(8)^j_i
+{{\gamma^*_{\overline{10}}}}\cdot(27^{\dagger})^i_j(27)^j_i .  
\label{eq:dec10b} 
\end{equation}
At this point it is a simple, but tedious and extensive, task to
expand this expression in terms of the tensor components. This is
easily done with the aid of a symbolic manipulator program.  The
result is shown as a table in  Sec.~\ref{10table}.
 
\subsection{$({\bf{27}},{\frac12})$ Decay}
\label{sec27}

The tensor decomposition in Eq.~(\ref{eq:decayprod}) contains a ${\bf{27}}$,
so  the strong decay amplitude of a
${\bf{27}}$ pentaquark already proceeds through  the flavor symmetric part
of the Hamiltonian. The matrix elements
\begin{equation}
\langle \mathbf{8} \otimes \mathbf{10} | O|\mathbf{27} \rangle
=\alpha^*_{27}\cdot\epsilon^{qjk}(M^{\dagger})^i_j(D^{\dagger})_{klr}
W^{lr}_{iq}, 
\label{eq:27singlet}
\end{equation}
where $|\mathbf{27} \rangle$ stands for a pentaquark state, are all
proportional to the single constant $\alpha^*_{27}$ and listed in
Sec.~\ref{27table}. The components of the tensor $W^{lr}_{iq}$ are
given in~\cite{paper}.

Turning to the contributions to the decay amplitude at first order in $SU(3)$
symmetry breaking, we analyze the generic matrix element
\begin{equation}
\langle \mathbf{8} \otimes \mathbf{10} | O^i_j(T_8)^j_i|\mathbf{27}
\rangle. 
\label{eq:matelement2}
\end{equation}
The  tensor decompositions
\begin{equation}
{\bf{8}}  \otimes {\bf{27}} = {\bf{8}} \oplus{\bf{10}}\oplus \overline{{\bf{10}}} \oplus {\bf{27}} \oplus {\bf{27}}\oplus {\bf{35}} \oplus \overline{{\bf{35}}} \oplus {\bf{64}}\label{eq:8x27}
\end{equation}
and Eq.~(\ref{eq:decayprod}) give that this matrix element is
parameterized by five reduced matrix elements:
\begin{multline}
\langle \mathbf{8} \otimes \mathbf{10} | O^i_j(T_8)^j_i |\mathbf{27}
\rangle=\beta^*_{27}\cdot(8^{\dagger})^i_j(8)^j_i+\gamma^*_{27}\cdot(10^{\dagger})_{ijk}(10)^{ijk}
+\delta^*_{27}\cdot(27)^i_j(27a)^j_i\\
+\epsilon^*_{27}\cdot(27^{\dagger})^i_j(27b)^j_i+\zeta^*_{27}(35^{\dagger})^m_{ijkl}(35)^{ijkl}_m.
\label{eq:dec27}
\end{multline}
The relevant tensors in this decomposition coming from
meson$\times$baryon states are given by Eq.~(\ref{eq:irrep1}) and
\begin{eqnarray}
&& (10^{\dagger})_{ijk}=(D^{\dagger})_{r(ij}(M^{\dagger})^r_{k)} \nonumber\\
&& (35^{\dagger})^{m}_{ijkl}=(D^{\dagger})_{(ijk}(M^{\dagger})^m_{l)}-\frac{1}{5}\delta^m_{(l}(10^{\dagger})_{ijk)} .
\label{eq:irrep3}
\end{eqnarray}
Their ${\bf{27}}\times (T_8)$ counterparts are composed of the $T_8$
Gell-Mann matrix and the tensor $W^{ij}_{kl}$:
\begin{eqnarray}
&& (8)^i_j=(T_8)^r_sW^{si}_{rj} \nonumber \\
&& (10)^{ijk}=\epsilon^{rs(i}W^{jk)}_{rq}(T_8)^q_s \nonumber \\
&& (27a)^{ik}_{jl}=(T_8)^r_{(j}W^{ik}_{l)r}-\frac{1}{5}\delta^{(i}_{(j}(8)^{k)}_{l)} \nonumber \\
&& (27b)^{ik}_{jl}=(T_8)^{(i}_rW^{k)r}_{jl}-\frac{1}{5}\delta^{(i}_{(j}(8)^{k)}_{l)} \nonumber \\
&& (35)^{ijkl}_{m}=(T_8)^{(i}_qW^{jk}_{mr}\epsilon^{l)qr}-\frac{1}{6}\delta^{(i}_m(10)^{jkl)}.\label{eq:irrep4}
\end{eqnarray}
All tensors correspond to irreducible representations of $SU(3)$ and
therefore are symmetric and traceless. The amplitudes
for specific decay modes that follow from Eq.~(\ref{eq:dec27}) are listed in
Sec.~\ref{27table}.

\subsection{$({\bf{35}},{\frac32})$ Decay}\label{sec35}

The tensor decomposition Eq.~(\ref{eq:decayprod}) contains a
${\mathbf{35}}$, so, as was the case with the ${\mathbf{27}}$, the decay
of the ${\mathbf{35}}$ pentaquark into a baryon decuplet state and a
pseudoscalar meson proceeds through the flavor symmetric part of the
Hamiltonian. Contrast this with the decay into a baryon octet and a
pseudoscalar meson, which is forbidden in the $SU(3)$ symmetry
limit~\cite{paper}. The values of matrix elements
\begin{equation}
\langle \mathbf{8} \otimes \mathbf{10} | O|\mathbf{35} \rangle=\alpha^*_{35}\cdot(M^{\dagger})^m_i(D^{\dagger})_{jkl} F^{ijkl}_m, \label{eq:35singlet}
\end{equation}
where $|\mathbf{35} \rangle$ stands for a pentaquark state, are all
proportional to the single parameter $\alpha^*_{35}$ and listed in the
table in Sec.~\ref{35table}. The components of the tensor $F^{ijkl}_m$
can be found in~\cite{paper}.

It follows from tensor decomposition
\begin{equation}
{\bf{35}}\otimes{\bf{8}}={\bf{10}}\oplus{\bf{27}}\oplus{\bf{28}}\oplus{\bf{35}}\oplus{\bf{35}}\oplus{\bf{64}}\oplus{\bf{81}}\label{eq:35x8}
\end{equation}
that the flavor symmetry breaking part of the decay amplitude is
parameterized by four reduced matrix elements:
\begin{multline}
\langle \mathbf{8} \otimes \mathbf{10} | O^i_j(T_8)^j_i |\mathbf{35} \rangle
=\beta^*_{35}\cdot(10^{\dagger})_{ijk}(10)^{ijk}+\gamma^*_{35}\cdot(27)^i_j(27)^j_i\\
+\delta^*_{27}\cdot(35^{\dagger})^m_{ijkl}(35_a)^{ijkl}_m+\epsilon^*_{27}\cdot(35^{\dagger})^m_{ijkl}(35_b)^{ijkl}_m.
\label{eq:dec35}
\end{multline}
The relevant tensors in the decomposition of ${\bf{35}}\times (T_8)$ are:
\begin{eqnarray}
&&(10)^{ijk}=(T_8)^r_sF^{sijk}_r \nonumber \\
&&(27)^{ik}_{jl}=\epsilon_{rs(j}F^{rqik}_{l)}(T_8)^s_q \nonumber \\
&&(35a)^{ijkl}_m=(T_8)^r_mF^{ijkl}_r-\frac{1}{6}\delta^{(i}_m(10)^{jkl)} \nonumber \\
&&(35b)^{ijkl}_m=(T_8)^{(i}_rF^{jkl)r}_m-\frac{1}{6}\delta^{(i}_m(10)^{jkl)}.\label{eq:irrep5}
\end{eqnarray}
The amplitudes for the specific decay modes in Eq.~(\ref{eq:dec35}) are listed
in Sec.~\ref{35table}.

\section{$SU(6)$ Calculation}
\label{su6}
In the previous section we saw how the decay amplitudes of pentaquarks
to a baryon decuplet are given in terms of a few reduced matrix
elements. The decays of pentaquarks into a baryon octet are similarly
given in terms of a different set of reduced matrix
elements. $SU(3)$-flavor symmetry gives no relation among the decay
parameters for the decuplet and octet final states. However, $SU(6)$
spin-flavor symmetry does relate the matrix elements of decays into
octet and decuplet final states. 

Spin-flavor $SU(6)$   is an approximate symmetry of  baryons. It was
motivated by the success of the non-relativistic quark model, in which
the $SU(3)$ of flavor and the separate $SU(2)$ symmetry of spin (an
internal symmetry in the non-relativistic limit) are immediately
promoted to a full $SU(6)$, and is supported empirically. More
recently it has been shown that, given some mild technical
assumptions,  contracted $SU(6)$ is a symmetry of
the baryon sector of QCD in the large $N_c$
limit\cite{dashen,largeN}, and that the pattern of symmetry breaking
implied by the large-$N_c$ works well\cite{jenkins}. While $SU(6)$
works relatively well for strong decays, its predictions are poor for
non-leptonic weak decays\cite{georgi}.

The baryon octet $(\mathbf{8},{\frac12})$ and decuplet
$(\mathbf{10},{\frac32})$ states comprise the $56$ multiplet
of $SU(6)$.  Hence pentaquark decays into the two multiplets are
related in the $SU(6)$ limit. Moreover, the
$(\overline{1},1,1,1,1)_{700}$ representation\footnote{We use the
common notation for irreducible representation of $SU(N)$:
$(n_1,n_2\ldots,n_k)$, where $n_i$ is the number of boxes in the
first, second etc. column of the Young tableaux starting from the
left. The $\overline{n}$ is equal to $N-n$ which corresponds to the fundamental
representation that is complex conjugate to $n$. The subscript shows
the dimension of the representation. Where it doesn't result in
ambiguity we use only dimension to identify a representation.} of
$SU(6)$ contains, among others, the
$(\mathbf{\overline{10}},{\frac12})$, $(\mathbf{27},{\frac12})$, and
$(\mathbf{35},{\frac32})$ representations. This suggests that
pentaquark multiplets exist not just in one of the representations
discussed in Sec.~\ref{su3} but rather in all, with the appropriate
spin\cite{penta}. In the following we assume that pentaquarks are
members of this $700$ representation, so we will be able to
relate the reduced matrix elements of the
$(\mathbf{\overline{10}},{\frac12})$, $(\mathbf{27},{\frac12})$, and
$(\mathbf{35},{\frac32})$ representations. The reader should bear in
mind, however, that individually each of these $SU(3)$ representations
may fit into other $SU(6)$ representation and have, possibly,
different spin, in which case there are no relations among decays of
the $\mathbf{\overline{10}}$, $\mathbf{27}$,
and $\mathbf{35}$ representations.

The $SU(6)$ calculation proceeds much as the $SU(3)$
calculation. Following~\cite{georgi} one builds $SU(6)$ tensors as a
direct product of irreducible $SU(2)$-spin and $SU(3)$-flavor
tensors. The $SU(2)$-spin tensors for spin ${\frac12}$ and ${\frac32}$
representations are
\begin{eqnarray}
& \chi^1=|{\frac12},{\frac12}\rangle \qquad \chi^2=|{\frac12},-{\frac12}\rangle &\nonumber \\
&\chi^{111}=|{\frac32},{\frac32}\rangle, \quad \chi^{112}={\frac1{\sqrt{3}}}|{\frac32},{\frac12}\rangle, \quad
\chi^{122}={\frac1{\sqrt{3}}}|{\frac32},-{\frac12}\rangle, \quad \chi^{222}=|{\frac32},-{\frac32}\rangle. &
\label{eq:spintensors}
\end{eqnarray}
The $56$-multiplet of baryonic states is represented by the
tensor~\cite{georgi}
\begin{equation}
(D^\dagger)_{aibjck}=\chi^{\dagger}_{ijk}(D^\dagger)_{abc}+{\frac1{3\sqrt{2}}}\left[\epsilon_{ij}\chi^{\dagger}_k\epsilon_{abd}(B^{\dagger})^d_c+\epsilon_{jk}\chi^{\dagger}_i\epsilon_{bcd}(B^{\dagger})^d_a+\epsilon_{ki}\chi^{\dagger}_j\epsilon_{cad}(B^{\dagger})^d_b \right].
\label{eq:56}
\end{equation}
Here $a,b,c=1,2,3$ are $SU(3)$ indices and $i,j=1,2$ are $SU(2)$
indices; $\epsilon_{ij}$ and $\epsilon_{abc}$ are invariant completely
antisymmetric $SU(2)$ and $SU(3)$ tensors. The  factor
${\frac1{3\sqrt{2}}}$ ensures the correct normalization.
The tensor in Eq.~(\ref{eq:56}) is completely symmetric under permutation of
the combined $SU(6)$ indices $(ai),(bj),(ck)$. Again we need complex
conjugate tensors because~(\ref{eq:56}) represents a $bra$ vector that
transforms as a complex conjugate representation of $SU(6)$. Hermitian
conjugation for $SU(2)$ tensors is performed by lowering the indices
with  $\epsilon_{ij}$:
\begin{equation}
\chi^{\dagger}_q=\epsilon_{qr}\chi^r \qquad \chi^{\dagger}_{ijk}=\epsilon_{ir}\epsilon_{js}\epsilon_{kt}\chi^{rst}.
\label{eq:conjugate}
\end{equation}
 
The tensor $(\bar{1},1,1,1,1)_{700}$ of $SU(6)$ is completely symmetric in
its four upper (quark) indices $(ai),(bj),(ck),(dl)$ and has one lower
(antiquark) index $(ft)$. It vanishes if any upper index is
contracted with the lower one. These properties unambiguously (up to an
overall normalization factor) define the embedding of
$(\mathbf{\overline{10}},{\frac12})$ of $SU(3)\times SU(2)$ into $700$
of $SU(6)$ by
\begin{eqnarray}
P(\mathbf{\overline{10}})^{aibjckdl}_{ft}&=&P_{fgh}\chi_t\left[\epsilon^{gab}\epsilon^{hcd}\epsilon^{ij}\epsilon^{kl}+\epsilon^{gac}\epsilon^{hbd}\epsilon^{ik}\epsilon^{jl}+\epsilon^{gad}\epsilon^{hbc}\epsilon^{il}\epsilon^{jk}\right]\nonumber \\
&-&{\frac16}\delta^a_f\delta^i_t P_{sgh}\chi_r\left[\epsilon^{gsb}\epsilon^{hcd}\epsilon^{rj}\epsilon^{kl}+\epsilon^{gsc}\epsilon^{hbd}\epsilon^{rk}\epsilon^{jl}+\epsilon^{gsd}\epsilon^{hbc}\epsilon^{rl}\epsilon^{jk}\right].
\label{eq:penta10b}
\end{eqnarray}
The embedding of $(\mathbf{27},{\frac12})$ and
$(\mathbf{35},{\frac32})$ into $700$ is even simpler and is given by
\begin{eqnarray}
&&P(\mathbf{27})^{aibjckdl}_{ft}
=W^{ab}_{fg}\epsilon^{gcd}\epsilon^{kl}(\chi^i\delta^j_t+\chi^j\delta^i_t)
+{\rm permutations} \nonumber \\
&&P(\mathbf{35})^{aibjckdl}_{ft}=F^{abcd}_f \chi^{(ijk}\delta^{l)}_t.
\label{eq:penta27&35}
\end{eqnarray}
In order to have  properly normalized tensors (so that each state
occurs with unit weight in
$(P^\dagger)_{aibjckdl}^{ft}P^{aibjckdl}_{ft} $), we write the $700$
decomposition in terms of the tensors in Eqs.~(\ref{eq:penta10b}) and~(\ref{eq:penta27&35}) as
\begin{equation}
\label{eq:700}
P^{aibjckdl}_{ft}=\frac{1}{6\sqrt{2}}P(\mathbf{\overline{10}})^{aibjckdl}_{ft}
+\frac{1}{12\sqrt{2}}P(\mathbf{27})^{aibjckdl}_{ft}
+\frac{1}{2\sqrt{5}}P(\mathbf{35})^{aibjckdl}_{ft}.
\end{equation}

A consistent way of including pseudoscalar mesons into the $SU(6)$
calculation is by the use of the axial current as an interpolating
field which transforms as $(\mathbf{8},1)$ under $SU(3)\times
SU(2)$~\cite{georgi}.  This is a part of the decomposition of
$(\overline{1},1)_{35}$ of $SU(6)$. So, {\it the $SU(6)$ calculation
is essentially chiral perturbation theory calculation}, where
pseudoscalar mesons are components of axial current of broken chiral
symmetry. Therefore when using $SU(6)$ relations for the decay
amplitudes one should multiply the amplitude by the momentum of meson
in the final state. In the limit of unbroken $SU(6)$ the momentum is
the same for all mesons. Following Ref.~\cite{georgi} the $SU(6)$ tensor for
the pseudoscalar octet is written in terms of the standard octet
tensor $M^a_b$ as
\begin{equation}
\label{eq:su6meson}
(M^{\dagger})^{ai}_{bj}
=(M^{\dagger})^a_b((\vec{s}\cdot\hat{n})^{\dagger})^i_j, 
\qquad {\rm where} \qquad (\vec{s}\cdot\hat{n})^i_j
=\left(\begin{array}{cc} \cos\theta & \sin\theta e^{-i\phi} \\ 
\sin\theta e^{i\phi} & - \cos\theta \end{array}\right)
\end{equation}
and $\hat{n}$ is an arbitrary unit vector denoting the direction of
the momentum of the meson. Of course, our symmetry relations~(\ref{eq:su6relation}) and~(\ref{eq:masterequation}) are independent of this direction.

In the limit of unbroken $SU(6)$ the decay modes
\begin{eqnarray}
&&(\overline{{\bf{10}}},{\frac12})\to({\bf{8}},{\frac12})+{\rm mesons}  \nonumber \\
&&({\bf{27}},{\frac12})\to({\bf{8}},{\frac12})+{\rm mesons}\qquad {\rm and}\qquad ({\bf{27}},{\frac12})\to({\bf{10}},{\frac32})+{\rm mesons} \nonumber \\
&&({\bf{35}},{\frac32})\to({\bf{10}},{\frac32})+{\rm mesons}\label{eq:decmodes}
\end{eqnarray}
are incorporated into the single matrix element
\begin{equation}
\langle 35 \otimes 56 |O|700 \rangle, \label{eq:matelement3}
\end{equation}
where $O$ is the $SU(6)$ invariant part of the effective decay
Hamiltonian and therefore are all proportional to a single
parameter\footnote{The resulting amplitudes are products of $SU(2)$ spin
amplitudes~(\ref{eq:spin1}) and~(\ref{eq:spin2}) and $SU(3)$ flavor amplitudes.}.  This gives the following
relations between the $SU(3)$ reduced matrix elements
$\alpha_{\overline{10}}$ and $\alpha_{27}$ from~\cite{paper} and
$\alpha^*_{27}$ and $\alpha^*_{35}$ from Secs.~\ref{27table}
and~\ref{35table}:
\begin{equation}
\label{eq:su6relation}
\alpha_{27}=\frac{\sqrt2}3\, \alpha_{\overline{10}},\qquad\qquad 
\alpha^*_{27}=\sqrt{\frac{2}{3}}\,\alpha_{\overline{10}}\qquad{\rm and}\qquad 
\alpha^*_{35}=\sqrt{2}\, \alpha_{\overline{10}}.
\end{equation}

The transitions
\begin{equation}
(\overline{{\bf{10}}},{\frac12})\to({\bf{10}},{\frac12})+{\rm mesons} \qquad {\rm and} \qquad ({\bf{35}},{\frac32})\to({\bf{8}},{\frac12})+{\rm mesons}\label{eq:forbidden}
\end{equation}
are forbidden in the limit of exact $SU(6)$ symmetry. The
corresponding amplitudes are due to flavor symmetry breaking terms in
the effective decay Hamiltonian. The simplest $SU(6)$ generalization
of the flavor symmetry breaking operator $T_8$ of $SU(3)_F$ is
obtained by combining it with $\delta^i_j$ of $SU(2)_{spin}$:
\begin{equation}
(T_8)^{ai}_{bj}=(T_8)^a_b\delta^i_j.
\label{eq:T8}
\end{equation}
This operator belongs to the $(\mathbf{8},0)$ representation of
$SU(3)\times SU(2)$ and therefore transforms as a $35$ under $SU(6)$ as
well. $SU(3)\times SU(2)$ decomposition of $(\bar{1},1)_{35}$ of $SU(6)$
contains also $({\bf 8},1)$ and $({\bf1},1)$. The corresponding
operators are $1/N_c$ suppressed and are not included in our analysis.

The tensor decomposition of products of $SU(6)$ representations 
\begin{eqnarray}
&& (\overline{1},1)_{35}\otimes(1,1,1)_{56}=(\overline{1},2,1,1)_{1134}\oplus(\overline{1},1,1,1,1)_{700}\oplus(2,1)_{70}\oplus(1,1,1)_{56} \nonumber \\
&&(\overline{1},1)_{35}\otimes(\overline{1},1,1,1,1)_{700}=(\overline{1},\overline{1},2,1,1,1)_{8624}\oplus(\overline{2},2,1,1,1)_{5670}\oplus(\overline{1},\overline{1},1,1,1,1,1)_{4536} \nonumber \\
&&~ \oplus(\overline{2},1,1,1,1,1)_{3080}\oplus(\overline{1},2,1,1)_{1134}\oplus(\overline{1},1,1,1,1)_{700}\oplus(\overline{1},1,1,1,1)_{700}\oplus(1,1,1)_{56}.
\end{eqnarray}
shows that there are four reduced matrix elements contributing to the
decay amplitude:
\begin{equation}
\langle 35 \otimes 56 | O^{ai}_{bj}(T_8)^{bj}_{ai}|700 \rangle =56\cdot56\oplus700\cdot700\oplus700\cdot700\oplus1134\cdot1134.
\label{eq:su6matelement}
\end{equation}
The corresponding irreducible tensors in the decomposition of
$35\otimes56$, the product of the final baryon and meson states, are
(here $A,B,C\ldots=1,\ldots6$ are $SU(6)$ indices):
\begin{eqnarray}
&& (\overline{56})_{ABC}=(D^{\dagger})_{(ABS}(M^{\dagger})^S_{C)} \nonumber \\
&& (\overline{700})^F_{ABCD}=(D^{\dagger})_{(ABC}(M^{\dagger})^F_{D)}-{\frac19}(\overline{56})_{(ABC}\delta^F_{D)} \nonumber \\
&& (\overline{1134})^F_{ABCD}=(D^{\dagger})_{ABC}(M^{\dagger})^F_D-(D^{\dagger})_{DBC}(M^{\dagger})^F_A \nonumber \\
&&+{\frac1{40}}\left( (D^{\dagger})_{ABS}(M^{\dagger})^S_C\delta^F_D+(D^{\dagger})_{ACS}(M^{\dagger})^S_B\delta^F_D-(D^{\dagger})_{DBS}(M^{\dagger})^S_C\delta^F_A-(D^{\dagger})_{CDS}(M^{\dagger})^S_B\delta^F_A\right) \nonumber \\
&&+{\frac18}\left((D^{\dagger})_{CDS}(M^{\dagger})^S_A\delta^F_B+(D^{\dagger})_{BDS}(M^{\dagger})^S_A\delta^F_C-(D^{\dagger})_{ABS}(M^{\dagger})^S_D\delta^F_C-(D^{\dagger})_{ACS}(M^{\dagger})^S_D\delta^F_B\right) \nonumber \\
&&+{\frac3{20}}\left((D^{\dagger})_{CBS}(M^{\dagger})^S_A\delta^F_D-(D^{\dagger})_{BCS}(M^{\dagger})^S_D\delta^F_A\right).
\label{eq:35x56}
\end{eqnarray}
We need hermitian conjugate tensors because they represent $bra$
vectors. The $\overline{56}$ is completely symmetric under permutation
of $A,B,C$. The $\overline{700}$ is completely symmetric under
permutation of $A,B,C,D$ and vanishes if its upper index is contracted
with a lower one. The $1134$ has  mixed symmetry: the indices $A,B,C$ are
symmetrized, while  $A,D$ are antisymmetrized. This fixes the first two
terms of $1134$. "The tail" is determined by tracelessness of $1134$,
i.e., the tensor must vanish if its upper index is contracted with a
lower one.

The relevant irreducible tensors in the decomposition of
$35\otimes700$, the product of the pentaquark and the symmetry
breaking Hamiltonian,  are:
\begin{eqnarray}
\label{eq:35x700}
&& (56)^{ABC}=(T_8)^R_SP^{ABCS}_R \nonumber \\
&& (700a)^{ABCD}_F=(T_8)^S_FP^{ABCD}_S
-{\frac19}(56)^{(ABC}\delta^{D)}_F \nonumber \\
&& (700b)^{ABCD}_F=(T_8)^{(D}_SP^{ABC)S}_F
-{\frac19}(56)^{(ABC}\delta^{D)}_F \nonumber \\
&& (1134)^{ABCD}_F=(T_8)^{D}_SP^{ABCS}_F-(T_8)^{A}_SP^{DBCS}_F
-{\frac15}\left[(56)^{ABC}\delta^D_F-(56)^{DBC}\delta^A_F\right].
\end{eqnarray}
These tensors satisfy the same symmetry requirements as those
in Eq.~(\ref{eq:35x56}).

At first order in $SU(3)$ breaking the amplitudes of a
$\mathbf{\overline{10}}$ pentaquark decay into baryon octet plus
pseudoscalar mesons are given in terms of four reduced matrix
elements\cite{paper}, $\beta_{\overline{10}}$,
$\gamma_{\overline{10}}$, $\delta_{\overline{10}}$, and
$\epsilon_{\overline{10}}$, the same number as for the $SU(6)$ decay
amplitudes calculated in this section. Matching between the $SU(3)$
and $SU(6)$ calculations allows one to express {\bf all} the reduced
matrix elements in the present paper and~\cite{paper} in terms of
these four parameters only:
\begin{eqnarray}
\label{su6results}
&&{{}\beta^*_{\overline{10}}}=
          \frac{4}{15}\left(-10\gamma_{\overline{10}}
	  +5\delta_{\overline{10}}
	  +2\sqrt{3}\epsilon_{\overline{10}}\right),
	  \nonumber \\
&&{{\gamma^*_{\overline{10}}}}=0, 
	  \nonumber \\
&&\beta_{35}=
	  -\frac{1}{\sqrt{3}}\left(-10\gamma_{\overline{10}}
	  +5\delta_{\overline{10}}+
	  2\sqrt{3}\epsilon_{\overline{10}}\right),
	  \nonumber \\
&& \gamma_{35}=0, \nonumber \\
&&\beta^*_{35}=-\frac5{\sqrt6}\,\gamma_{\overline{10}}
               +\sqrt{\frac32}~\delta_{\overline{10}}
               +\frac{17}{15\sqrt{2}}\,\epsilon_{\overline{10}},
               \nonumber \\      
&&\gamma^*_{35}=\frac1{4\sqrt2}\,\epsilon_{\overline{10}},\nonumber \\
&& \delta^*_{35}=-\frac1{2\sqrt6}\,\beta_{\overline{10}}
+\frac5{3\sqrt6}\,\gamma_{\overline{10}}
               -\frac1{3\sqrt6}\,\delta_{\overline{10}}
               -\frac{11}{30\sqrt{2}}\,\epsilon_{\overline{10}},
                                                      \nonumber \\ 
&&  \epsilon^*_{35}=\frac1{2\sqrt6}\,\beta_{\overline{10}}
+\frac5{6\sqrt6}\,\gamma_{\overline{10}}
               -\frac1{6\sqrt6}\,\delta_{\overline{10}}
               -\frac{11}{60\sqrt{2}}\,\epsilon_{\overline{10}},
	       \nonumber \\ 
&& \beta_{27}=
        -\frac{14}{9\sqrt{5}}\,\gamma_{\overline{10}}
        +\frac{32}{45\sqrt{5}}\,\delta_{\overline{10}}
        +\frac{128}{225\sqrt{15}}\,\epsilon_{\overline{10}},
        \nonumber \\
&&\delta_{27}=-\frac{5}{6\sqrt{3}}\,\gamma_{\overline{10}}
           +\frac{1}{6\sqrt{3}}\,\delta_{\overline{10}}
           +\frac{7}{45}\,\epsilon_{\overline{10}},
    \nonumber \\
&&\gamma_{27}=
      -\frac{52}{9\sqrt{5}}\,\gamma_{\overline{10}}
      +\frac{142}{45\sqrt{5}}\,\delta_{\overline{10}}
       +\frac{808}{225\sqrt{15}}\,\epsilon_{\overline{10}},
        \nonumber \\
&&\epsilon_{27}=
          \frac{1}{9}\epsilon_{\overline{10}},
    \nonumber \\
&&\zeta_{27}=
        -\frac{1}{3\sqrt{3}}\,\beta_{\overline{10}}
        +\frac{5}{18\sqrt{3}}\,\gamma_{\overline{10}}
        -\frac{1}{18\sqrt{3}}\,\delta_{\overline{10}}
        -\frac{4}{45}\,\epsilon_{\overline{10}},
   \nonumber \\
&&\psi_{27}=
      \frac{1}{3\sqrt{3}}\,\beta_{\overline{10}}
      +\frac{5}{9\sqrt{3}}\,\gamma_{\overline{10}}
      -\frac{1}{9\sqrt{3}}\,\delta_{\overline{10}}
      -\frac{8}{45}\,\epsilon_{\overline{10}},
\nonumber \\
&&\beta^*_{27}=
        \frac{16\sqrt2}9~\gamma_{\overline{10}}
	-\frac{52\sqrt2}{45}~\delta_{\overline{10}}
	-\frac{248}{225}\sqrt{\frac23}~\epsilon_{\overline{10}},
      \nonumber \\
&&\delta^*_{27}=
          \frac{1}{3\sqrt{2}}\,\beta_{\overline{10}}
          -\frac{5}{18\sqrt2}\,\gamma_{\overline{10}}
          +\frac{1}{18\sqrt2}\,\delta_{\overline{10}}
          +\frac{1}{60\sqrt6}\,\epsilon_{\overline{10}},
         \nonumber \\
&&\gamma^*_{27}=
	 \frac{5}{27\sqrt{2}}\,\gamma_{\overline{10}}
	 -\frac{1}{27\sqrt{2}}\,\delta_{\overline{10}}
	 -\frac{19}{135\sqrt{6}}\,\epsilon_{\overline{10}},
         \nonumber \\
&&\epsilon^*_{27}=
       -\frac{1}{3\sqrt{2}}\,\beta_{\overline{10}}
       -\frac{5}{9\sqrt{2}}\,\gamma_{\overline{10}}
       +\frac{1}{9\sqrt{2}}\,\delta_{\overline{10}}
       +\frac{1}{30\sqrt{6}}\,\epsilon_{\overline{10}},
       \nonumber \\
&&\zeta^*_{27}=
       \frac{1}{36\sqrt{6}}\,\epsilon_{\overline{10}}.
\label{eq:masterequation}
\end{eqnarray}

 From the first two lines in Eq.~(\ref{eq:masterequation}) we see a
remarkably simple pattern. In the leading order all decay amplitudes
in~(\ref{eq:decmodes}) are proportional to the single
parameter~$\alpha_{\overline{10}}$. The decay modes
in~(\ref{eq:forbidden}), forbidden in the symmetry limit, are also
proportional to a single parameter, $(10\gamma_{\overline{10}}
-5\delta_{\overline{10}} -2\sqrt{3}\epsilon_{\overline{10}})$ at  first order in $m_s$. 

\section{Decay Amplitudes of $\overline{{\bf{10}}}$ into Decuplet
  Baryons}
\label{10table}

Decay amplitudes of $\overline{{\bf{10}}}$ pentaquarks to the baryon
decuplet are forbidden in the $SU(3)$ symmetry limit. Some decay modes
are also kinematically forbidden according to the reported values for pentaquark masses
$\Theta_{\overline{10}}^+\sim1540$~MeV~\cite{theta} and
$\Xi^{--}\sim~1862$~MeV~\cite{xi}. In the first order of $SU(3)$
breaking all isomultiplets in the $\overline{{\bf{10}}}$
representation are equidistant, so the multiplet masses, in MeV, are roughly
\begin{equation}
\Theta_{\overline{10}}^+\sim1540,\qquad 
N_{\overline{10}}\sim1647,\qquad 
\Sigma_{\overline{10}}\sim 1755, \qquad 
\Xi_{\overline{10}}\sim 1862.
\label{masses}
\end{equation}
It follows that the decay of a $\Theta^+$ into baryon decuplet states
is kinematically forbidden.  The only decay
modes of $\Theta^+_{\overline{10}} (0,+1)$ allowed by flavor
conservation are\footnote{The numbers in parentheses stand for
isospin and strangeness.} $\Delta^+(1/2,0)K^0(-1/2,+1)$ and
$\Delta^0(-1/2,0)K^+(1/2,0)$. These decays are forbidden:
1540MeV vs 1240+495=1735MeV. In fact, the only kinematically allowed
modes of members of the $\overline{\mathbf{10}}$ are $N_{\overline{10}} \to\Delta\pi$ ($|\vec p|
= 340$~MeV), $\Sigma_{\overline{10}} \to\Delta K$ ($|\vec p| = 140$~MeV),
and $\Xi_{\overline{10}} \to \Xi^*\pi $ ($|\vec p| = 270$~MeV).  

The table below shows the values of the amplitudes in terms of two
reduced $SU(3)$ matrix elements. We include in the table all modes
regardless of whether the process is kinematically allowed. The
amplitudes for kinematically forbidden modes may be useful in
computing off-shell amplitudes. 

\begin{center}
\begin{tabular}{|c|c|c|} \hline
$\Theta^+$ & ${{}\beta^*_{\overline{10}}}$ & ${{\gamma^*_{\overline{10}}}}$  \\ \hline
$\Delta^+K^0$ & 0 & 0 \\ \hline
$\Delta^0K^+$& 0 & 0 \\ \hline
\end{tabular}
\end{center}

\begin{center}
\begin{tabular}{|c|c|c|||c|c|c|} \hline
$N_{\overline{10}}$ & ${{}\beta^*_{\overline{10}}}$ & ${{\gamma^*_{\overline{10}}}}$ & $P_{\overline{10}}$ & ${{}\beta^*_{\overline{10}}}$ & ${{\gamma^*_{\overline{10}}}}$ \\ \hline
$\Delta^0\pi^0$ & $-1$ & $\sqrt{2}$ & $\Delta^+\pi^0$ & 1 & $-\sqrt{2}$ \\ \hline
$\Delta^{+}\pi^-$ & $\sqrt{1/2}$ & $-1$ & $\Delta^0\pi^+$ & $\sqrt{1/2}$ & $-1$ \\ \hline
$\Delta^-\pi^+$ & $-\sqrt{3/2}$ & $\sqrt{3}$ & $\Delta^{++}\pi^-$ & $-\sqrt{3/2}$ & $\sqrt{3}$\\ \hline
$\Sigma^{*0}K^0$ & $1/2$ & $2\sqrt{2}$ & $\Sigma^{*+}K^0$ & $-\sqrt{1/2}$ & $-4$\\ \hline
$\Sigma^{*-}K^+$ & $-\sqrt{1/2}$ & $-4$ & $\Sigma^{*0}K^+$ & $1/2$ & $2\sqrt{2}$\\ \hline
\end{tabular}
\end{center}

\begin{center}
\begin{tabular}{|c|c|c|||c|c|c|||c|c|c|} \hline
$\Sigma^-_{\overline{10}}$ & ${{}\beta^*_{\overline{10}}}$ & ${{\gamma^*_{\overline{10}}}}$ & $\Sigma^0_{\overline{10}}$ & ${{}\beta^*_{\overline{10}}}$ & ${{\gamma^*_{\overline{10}}}}$ & $\Sigma^+_{\overline{10}}$ & ${{}\beta^*_{\overline{10}}}$ & ${{\gamma^*_{\overline{10}}}}$  \\ \hline
$\Delta^-\overline{K^0}$ & $\sqrt{3/2}$ & $-\sqrt{3}$ & $\Delta^0\overline{K^0}$ & $-1$ & $\sqrt{2}$  & $\Delta^{++}K^-$ & $\sqrt{3/2}$ & $-\sqrt{3}$ \\ \hline
$\Delta^0K^-$ & $\sqrt{1/2}$ & $-1$ & $\Delta^+K^-$ & $-1$ & $\sqrt{2}$ &  $\Delta^+\overline{K^0}$ & $\sqrt{1/2}$ & $-1$  \\ \hline
$\Sigma^{*-}\pi^0$ & $1/2$ & $-3\sqrt{2}$ & $\Sigma^{*-}\pi^+$ & $1/2$ & $-3\sqrt{2}$ & $\Sigma^{*0}\pi^+$ & $-1/2$ & $3\sqrt{2}$ \\ \hline
$\Sigma^{*0}\pi^-$ & $-1/2$ & $3\sqrt{2}$ & $\Sigma^{*+}\pi^-$ & $1/2$ & $-3\sqrt{2}$ & $\Sigma^{*+}\pi^0$
& $-1/2$ & $3\sqrt{2}$ \\ \hline
$\Sigma^{*-}\eta$ & $-\sqrt{3}/2$ & $-2\sqrt{6}$ & $\Sigma^{*0}\eta$ & $\sqrt{3}/2$ & $2\sqrt{6}$ &  $\Sigma^{*+}\eta$ & $-\sqrt{3}/2$ & $-2\sqrt{6}$ \\ \hline
$\Xi^{*-}K^0$ & $-\sqrt{1/2}$ & $-4$ & $\Xi^{*-}K^+$ & $1/2$ & $2\sqrt{2}$ & $\Xi^{*0}K^+$ & $-\sqrt{1/2}$ & $-4$ \\ \hline
&&& $\Xi^{*0}K^0$ & $1/2$ & $2\sqrt{2}$&&&\\ \hline
\end{tabular}
\end{center}

\begin{center}
\begin{tabular}{|c|c|c|||c|c|c|||c|c|c|||c|c|c|} \hline
$\Xi^{--}_{\overline{10}}$ & ${{}\beta^*_{\overline{10}}}$ & ${{\gamma^*_{\overline{10}}}}$ &  $\Xi^{-}_{\overline{10}}$ & ${{}\beta^*_{\overline{10}}}$ & ${{\gamma^*_{\overline{10}}}}$ & $\Xi^{0}_{\overline{10}}$ & ${{}\beta^*_{\overline{10}}}$ & ${{\gamma^*_{\overline{10}}}}$ & $\Xi^{+}_{\overline{10}}$ & ${{}\beta^*_{\overline{10}}}$ & ${{\gamma^*_{\overline{10}}}}$ \\ \hline
$\Xi^{*-}\pi^-$ & $0$ & $-5\sqrt{3}$ & $\Xi^{*-}\pi^0$ & $0$ & $5\sqrt{2}$ & $\Xi^{*-}\pi^+$ & 0 & 5 & $\Xi^{*0}\pi^+$ & 0 & $-5\sqrt{3}$ \\ \hline
$\Sigma^{*-}K^-$ & $0$  & $5\sqrt{3}$  & $\Xi^{*0}\pi^-$ & $0$ & $5$ & $\Xi^{*0}\pi^0$ & 0 & $-5\sqrt{2}$ & $\Sigma^{*+}\overline{K^0}$ & $0$ & $5\sqrt{3}$\\ \hline
& & & $\Sigma^{*-}\overline{K^0}$ & $0$ & $5$ & $\Sigma^{*0}\overline{K^0}$ & $0$ & $-5\sqrt{2}$ & & &\\ \hline
& & & $\Sigma^{*0}K^-$ & $0$ & $-5\sqrt{2}$ & $\Sigma^{*+}K^-$& $0$ & $5$ & & & \\ \hline 
\end{tabular}
\end{center}

\section{Decay Amplitudes of ${\bf{27}}$ Pentaquark Multiplet}
\label{27table}
Below we list the decays of the twentyseven states that form this
irreducible representation of $SU(3)$ in terms of one parameter,
$\alpha^*_{27}$, at zeroth order in $SU(3)$ symmetry breaking, plus
five more parameters, $\beta^*_{27}$, $\gamma^*_{27}$,
$\delta^*_{27}$, $\epsilon^*_{27}$, and~$\zeta^*_{27}$ of order
$m_s$. It is worth mentioning that there are several new ``exotic
decays''. For example, the decays $\Sigma^{\prime++}_{27}\to
\Delta^{++}K_{S,L}$ and $\Sigma^{\prime-}_{27}\to \Delta^{-}K^-$ have
final state flavor $uuus\bar d$ (or $uuud\bar s$) and $ddds\bar u $,
respectively.
 
\begin{center}
\begin{tabular}{|c|c|c|c|c|c|c|} \hline
$\Theta^{++}_{27}$ & $\alpha^*_{27}$ & $\beta^*_{27}$ & $\gamma^*_{27}$ & $\delta^*_{27}$ & $\epsilon^*_{27}$& $\zeta^*_{27}$ \\ \hline
$\Delta^{++}K^0$ & $-1$ & $0$ & $0$ & $-8$ & $4$ &  $0$ \\ \hline
$\Delta^{+}K^+$ & $\sqrt{1/3}$ & $0$ & $0$ & $8/\sqrt{3}$ & $-4/\sqrt{3}$ & $0$ \\ \hline
\end{tabular}
\end{center}

\begin{center}
\begin{tabular}{|c|c|c|c|c|c|c|} \hline
$\Theta^+_{27}$ & $\alpha^*_{27}$ & $\beta^*_{27}$ & $\gamma^*_{27}$ & $\delta^*_{27}$ & $\epsilon^*_{27}$& $\zeta^*_{27}$ \\ \hline
$\Delta^+K^0$ & $-\sqrt{2/3}$ & $0$ & $0$ & $-8\sqrt{2/3}$ & $4\sqrt{2/3}$ & $0$ \\ \hline
$\Delta^0K^+$ & $\sqrt{2/3}$ & $0$ & $0$ & $8\sqrt{2/3}$ & $-4\sqrt{2/3}$ & $0$  \\ \hline
\end{tabular}
\end{center}

\begin{center}
\begin{tabular}{|c|c|c|c|c|c|c|} \hline
$\Theta^0_{27}$  & $\alpha^*_{27}$ & $\beta^*_{27}$ & $\gamma^*_{27}$ & $\delta^*_{27}$ & $\epsilon^*_{27}$& $\zeta^*_{27}$ \\ \hline
$\Delta^0K^0$ & $-\sqrt{1/3}$ & $0$ & $0$ & $-8/\sqrt{3}$ & $4/\sqrt{3}$ & $0$ \\ \hline
$\Delta^-K^+$  & $1$ & $0$ & $0$ & $8$ & $-4$ & $0$ \\ \hline
\end{tabular}
\end{center}

\begin{center}
\begin{tabular}{|c|c|c|c|c|c|c|} \hline
$N^+_{27}$ &  $\alpha^*_{27}$ & $\beta^*_{27}$ & $\gamma^*_{27}$ & $\delta^*_{27}$ & $\epsilon^*_{27}$& $\zeta^*_{27}$ \\ \hline
$\Delta^{++}\pi^-$ & $-\sqrt{2/15}$ & $3\sqrt{3/10}$ & $0$ & $-28/5\sqrt{2/15}$ & $2/5\sqrt{2/15}$ & $0$ \\ \hline
$\Delta^{+}\pi^0$ & $2/3\sqrt{5}$ & $-3/\sqrt{5}$ & $0$ & $56/15\sqrt{5}$ & $-4/15\sqrt{5}$ & $0$ \\ \hline
$\Delta^0\pi^+$ & $\sqrt{2}/3\sqrt{5}$ & $-3/\sqrt{10}$ & $0$ & $28/15\sqrt{2/5}$ & $-2/15\sqrt{2/5}$ & $0$ \\ \hline
$\Sigma^{*+}K^0$& $4/3\sqrt{2/5}$ & $3/\sqrt{10}$ & $0$ & $112/15\sqrt{2/5}$ & $-8/15\sqrt{2/5}$ & $0$ \\ \hline
$\Sigma^{*0}K^+$ & $-4/3\sqrt{5}$ & $-3/2\sqrt{5}$ & $0$ & $-112/15\sqrt{5}$ &  $8/15\sqrt{5}$ & $0$ \\ \hline
\end{tabular}
\end{center}

\begin{center}
\begin{tabular}{|c|c|c|c|c|c|c|} \hline
$N^0_{27}$ &  $\alpha^*_{27}$ & $\beta^*_{27}$ & $\gamma^*_{27}$ & $\delta^*_{27}$ & $\epsilon^*_{27}$& $\zeta^*_{27}$ \\ \hline
$\Delta^{+}\pi^-$ & $-1/3\sqrt{2/5}$ & $3/\sqrt{10}$ & $0$ & $-28/15\sqrt{2/5}$ & $2/15\sqrt{2/5}$ & $0$ \\ \hline
$\Delta^{0}\pi^0$ & $2/3\sqrt{5}$ & $-3/\sqrt{5}$ & $0$ & $56/15\sqrt{5}$ & $-4/15\sqrt{5}$ & $0$ \\ \hline
$\Delta^-\pi^+$ & $\sqrt{2/15}$ & $-3\sqrt{3/10}$ & $0$ & $28/5\sqrt{2/15}$ & $-2/5\sqrt{2/15}$ & $0$ \\ \hline
$\Sigma^{*0}K^0$& $4/3\sqrt{5}$ & $3/2\sqrt{5}$ & $0$ & $112/15\sqrt{5}$ & $-8/15\sqrt{5}$ & $0$ \\ \hline 
$\Sigma^{*-}K^+$ & $-4/3\sqrt{2/5}$ & $-3/\sqrt{10}$ & $0$ & $-112/15\sqrt{2/5}$ & $8/15\sqrt{2/5}$ & $0$ \\ \hline
 \end{tabular}
\end{center}

\begin{center} 
\begin{tabular}{|c|c|c|c|c|c|c|} \hline
$\Delta^{++}_{27}$ &  $\alpha^*_{27}$ & $\beta^*_{27}$ & $\gamma^*_{27}$ & $\delta^*_{27}$ & $\epsilon^*_{27}$& $\zeta^*_{27}$ \\ \hline
$\Delta^{++}\pi^0$ & $1/2$ & $0$ & $27/2$ & $1$ & $-2$ & $-9$ \\ \hline
$\Delta^{++}\eta$ & $-\sqrt{3}/2$ & $0$ & $9\sqrt{3}/2$ & $-\sqrt{3}$ & $2\sqrt{3}$ & $-15\sqrt{3}$ \\ \hline
$\Delta^{+}\pi^+$ & $\sqrt{1/6}$ & $0$ & $9\sqrt{3/2}$ & $\sqrt{2/3}$ & $-2\sqrt{2/3}$ & $-3\sqrt{6}$ \\ \hline
$\Sigma^{*+}K^+$ & $-\sqrt{1/6}$ &  $0$ & $9\sqrt{3/2}$ & $-\sqrt{2/3}$ & $2\sqrt{2/3}$ & $15\sqrt{6}$ \\ \hline
\end{tabular}
\end{center}

\begin{center}
\begin{tabular}{|c|c|c|c|c|c|c|c|c|} \hline
$\Delta^{+}_{27}$ &  $\alpha^*_{27}$ & $\beta^*_{27}$ & $\gamma^*_{27}$ & $\delta^*_{27}$ & $\epsilon^*_{27}$& $\zeta^*_{27}$ \\ \hline
$\Delta^{++}\pi^-$ & $-\sqrt{1/6}$ & $0$ & $-9\sqrt{3/2}$ & $-\sqrt{2/3}$ & $2\sqrt{2/3}$ & $3\sqrt{6}$ \\ \hline
$\Delta^+\pi^0$ & $-1/6$ & $0$ & $-9/2$ & $-1/3$ & $2/3$ & $3$ \\ \hline
$\Delta^+\eta$ &  $\sqrt{3}/2$ & $0$ & $-9\sqrt{3}/2$ & $\sqrt{3}$ & $-2\sqrt{3}$ & $15\sqrt{3}$ \\ \hline
$\Delta^0\pi^+$ & $-\sqrt{2}/3$ & $0$ & $-9\sqrt{2}$ & $-2\sqrt{2}/3$ & $4\sqrt{2}/3$ & $6\sqrt{2}$ \\ \hline
$\Sigma^{*+}K^0$ & $1/3\sqrt{2}$ & $0$ & $-9/\sqrt{2}$ & $\sqrt{2}/3$ & $-2\sqrt{2}/3$ & $-15\sqrt{2}$ \\ \hline
$\Sigma^{*0}K^+$ & $1/3$ & $0$ & $-9$ & $2/3$ & $-4/3$ & $-30$ \\ \hline
\end{tabular}
\end{center}

\begin{center}
\begin{tabular}{|c|c|c|c|c|c|c|c|c|} \hline
$\Delta^{0}_{27}$ & $\alpha^*_{27}$ & $\beta^*_{27}$ & $\gamma^*_{27}$ & $\delta^*_{27}$ & $\epsilon^*_{27}$& $\zeta^*_{27}$ \\ \hline
$\Delta^{+}\pi^-$ & $\sqrt{2}/3$ & $0$ & $9\sqrt{2}$ & $2\sqrt{2}/3$ & $-4\sqrt{2}/3$ & $-6\sqrt{2}$ \\ \hline
$\Delta^{0}\pi^0$ & $-1/6$ & $0$ & $-9/2$ & $-1/3$ & $2/3$ & $3$ \\ \hline
$\Delta^{0}\eta$ & $-\sqrt{3}/2$ & $0$ & $9\sqrt{3}/2$ & $-\sqrt{3}$ & $2\sqrt{3}$ & $-15\sqrt{3}$ \\ \hline
$\Delta^{-}\pi^+$ & $\sqrt{1/6}$ & $0$ & $9\sqrt{3/2}$ & $\sqrt{2/3}$ & $-2\sqrt{2/3}$ & $-3\sqrt{6}$ \\ \hline
$\Sigma^{*0}K^0$ & $-1/3$ & $0$ & $9$ & $-2/3$ & $4/3$ & $30$  \\ \hline
$\Sigma^{*-}K^+$ & $-1/3\sqrt{2}$ & $0$ & $9/\sqrt{2}$ & $-\sqrt{2}/3$ & $2\sqrt{2}/3$ & $15\sqrt{2}$ \\ \hline
\end{tabular}
\end{center}

\begin{center}
\begin{tabular}{|c|c|c|c|c|c|c|} \hline
$\Delta^{-}_{27}$ & $\alpha^*_{27}$ & $\beta^*_{27}$ & $\gamma^*_{27}$ & $\delta^*_{27}$ & $\epsilon^*_{27}$& $\zeta^*_{27}$ \\ \hline
$\Delta^0\pi^-$ & $-\sqrt{1/6}$ & $0$ & $-9\sqrt{3/2}$ & $-\sqrt{2/3}$ & $2\sqrt{2/3}$ & $3\sqrt{6}$ \\ \hline
$\Delta^-\pi^0$ & $1/2$ & $0$ & $27/2$ & $1$ & $-2$ & $-9$ \\ \hline
$\Delta^-\eta$ & $\sqrt{3}/2$ & $0$ & $-9\sqrt{3}/2$ & $\sqrt{3}$ & $-2\sqrt{3}$ & $15\sqrt{3}$ \\ \hline
$\Sigma^{*-}K^0$ & $\sqrt{1/6}$ & $0$ & $-9\sqrt{3/2}$ & $\sqrt{2/3}$  & $-2\sqrt{2/3}$ & $-15\sqrt{6}$ \\ \hline
\end{tabular}
\end{center}

\begin{center}
\begin{tabular}{|c|c|c|c|c|c|c|} \hline
$\Sigma^{'++}_{27}$ & $\alpha^*_{27}$ & $\beta^*_{27}$ & $\gamma^*_{27}$ & $\delta^*_{27}$ & $\epsilon^*_{27}$& $\zeta^*_{27}$ \\ \hline
 $\Delta^{++}\overline{K^0}$ & $1$ & $0$ & $0$ & $-4$ & $-4$ & $36$ \\ \hline
 $\Sigma^{*+}\pi^+$ & $-\sqrt{1/3}$ & $0$ & $0$ & $4/\sqrt{3}$ & $4/\sqrt{3}$ & $36\sqrt{3}$  \\ \hline
\end{tabular}
\end{center}

\begin{center}
\begin{tabular}{|c|c|c|c|c|c|c|} \hline
$\Sigma^{'+}_{27}$ & $\alpha^*_{27}$ & $\beta^*_{27}$ & $\gamma^*_{27}$ & $\delta^*_{27}$ & $\epsilon^*_{27}$& $\zeta^*_{27}$ \\ \hline
$\Delta^{++}K^-$ & $1/2$ & $0$ & $0$ & $-2$ & $-2$ & $18$ \\ \hline
$\Delta^{+}\overline{K^0}$ & $-\sqrt{3}/2$ & $0$ & $0$ & $2\sqrt{3}$ & $2\sqrt{3}$ & $-18\sqrt{3}$ \\ \hline
$\Sigma^{*+}\pi^0$ & $-\sqrt{1/6}$ & $0$ & $0$ & $2\sqrt{2/3}$ & $2\sqrt{2/3}$ & $18\sqrt{6}$ \\ \hline
$\Sigma^{*0}\pi^+$ & $\sqrt{1/6}$ & $0$ & $0$ & $-2\sqrt{2/3}$ & $-2\sqrt{2/3}$ & $-18\sqrt{6}$ \\ \hline
\end{tabular}
\end{center}

\begin{center}
\begin{tabular}{|c|c|c|c|c|c|c|} \hline
$\Sigma^{'0}_{27}$ & $\alpha^*_{27}$ & $\beta^*_{27}$ & $\gamma^*_{27}$ & $\delta^*_{27}$ & $\epsilon^*_{27}$& $\zeta^*_{27}$ \\ \hline
$\Delta^+K^-$ & $-\sqrt{1/2}$ & $0$ & $0$ & $2\sqrt{2}$ & $2\sqrt{2}$ & $-18\sqrt{2}$ \\ \hline
$\Delta^0\overline{K^0}$ & $\sqrt{1/2}$ & $0$ & $0$ & $-2\sqrt{2}$ & $-2\sqrt{2}$ & $18\sqrt{2}$ \\ \hline 
$\Sigma^{*+}\pi^-$ & $1/3\sqrt{2}$ & $0$ & $0$ & $-2\sqrt{2}/3$ & $-2\sqrt{2}/3$ & $-18\sqrt{2}$ \\ \hline
$\Sigma^{*0}\pi^0$ & $\sqrt{2}/3$ & $0$ & $0$ & $-4\sqrt{2}/3$ & $-4\sqrt{2}/3$ & $-36\sqrt{2}$ \\ \hline
$\Sigma^{*-}\pi^+$ & $-1/3\sqrt{2}$ & $0$ & $0$ & $2\sqrt{2}/3$ & $2\sqrt{2}/3$ & $18\sqrt{2}$ \\ \hline
\end{tabular}
\end{center}

\begin{center}
\begin{tabular}{|c|c|c|c|c|c|c|} \hline
$\Sigma^{'-}_{27}$ & $\alpha^*_{27}$ & $\beta^*_{27}$ & $\gamma^*_{27}$ & $\delta^*_{27}$ & $\epsilon^*_{27}$& $\zeta^*_{27}$ \\ \hline
$\Delta^{0}K^-$ & $-\sqrt{3}/2$ & $0$ & $0$ & $2\sqrt{3}$ & $2\sqrt{3}$ & $-18\sqrt{3}$ \\ \hline
$\Delta^{-}\overline{K^0}$ & $1/2$ & $0$ & $0$ & $-2$ & $-2$ & $18$ \\ \hline
$\Sigma^{*0}\pi^-$  & $\sqrt{1/6}$ & $0$ & $0$ & $-2\sqrt{2/3}$ & $-2\sqrt{2/3}$ & $-18\sqrt{6}$ \\ \hline
$\Sigma^{*-}\pi^0$ & $\sqrt{1/6}$ & $0$ & $0$ & $-2\sqrt{2/3}$ & $-2\sqrt{2/3}$ & $-18\sqrt{6}$ \\ \hline
\end{tabular}
\end{center}

\begin{center}
\begin{tabular}{|c|c|c|c|c|c|c|} \hline
$\Sigma^{'--}_{27}$ & $\alpha^*_{27}$ & $\beta^*_{27}$ & $\gamma^*_{27}$ & $\delta^*_{27}$ & $\epsilon^*_{27}$& $\zeta^*_{27}$ \\ \hline
$\Delta^{-}K^-$ &  $-1$ & $0$ & $0$ & $4$ & $4$ & $-36$ \\ \hline
$\Sigma^{*-}\pi^-$ & $\sqrt{1/3}$ & $0$ & $0$ & $-4/\sqrt{3}$ & $-4/\sqrt{3}$ & $-36\sqrt{3}$ \\ \hline
\end{tabular}
\end{center}

\begin{center}
\begin{tabular}{|c|c|c|c|c|c|c|} \hline
$\Sigma^+_{27}$ & $\alpha^*_{27}$ & $\beta^*_{27}$ & $\gamma^*_{27}$ & $\delta^*_{27}$ & $\epsilon^*_{27}$& $\zeta^*_{27}$ \\ \hline
$\Delta^{++}K^-$ & $1/2\sqrt{5}$ & $-3/\sqrt{5}$ & $-18/\sqrt{5}$ & $2/5\sqrt{5}$ & $2/5\sqrt{5}$ & $6\sqrt{5}$ \\ \hline
$\Delta^{+}\overline{K^0}$ & $1/2\sqrt{15}$ & $-\sqrt{3/5}$ & $-6\sqrt{3/5}$ & $2/5\sqrt{15}$ & $2/5\sqrt{15}$ & $2\sqrt{15}$ \\ \hline
$\Sigma^{*+}\pi^0$ & $-\sqrt{3/10}$ & $\sqrt{3/10}$ & $-6\sqrt{6/5}$ & $-2/5\sqrt{6/5}$ & $-2/5\sqrt{6/5}$ & $2\sqrt{30}$ \\ \hline
$\Sigma^{*+}\eta$ & $\sqrt{2/5}$ & $3/\sqrt{10}$ & $0$ & $4/5\sqrt{2/5}$ & $4/5\sqrt{2/5}$ & $12\sqrt{10}$ \\ \hline
$\Sigma^{*0}\pi^+$ & $-\sqrt{3/10}$ & $\sqrt{3/10}$ & $-6\sqrt{6/5}$ & $-2/5\sqrt{6/5}$ & $-2/5\sqrt{6/5}$ & $2\sqrt{30}$ \\ \hline
$\Xi^{0}K^+$ & $2/\sqrt{15}$ & $\sqrt{3/5}$ & $-12\sqrt{3/5}$ & $8/5\sqrt{15}$ & $8/5\sqrt{15}$ & $-8\sqrt{15}$ \\ \hline

\end{tabular}
\end{center}

\begin{center}
\begin{tabular}{|c|c|c|c|c|c|c|} \hline
$\Sigma^0_{27}$ & $\alpha^*_{27}$ & $\beta^*_{27}$ & $\gamma^*_{27}$ & $\delta^*_{27}$ & $\epsilon^*_{27}$& $\zeta^*_{27}$ \\ \hline
$\Delta^+K^-$ & $\sqrt{1/30}$ & $-\sqrt{6/5}$ & $-6\sqrt{6/5}$ & $2/5\sqrt{2/15}$ & $2/5\sqrt{2/15}$ & $2\sqrt{30}$ \\ \hline
$\Delta^0\overline{K^0}$ & $\sqrt{1/30}$ & $-\sqrt{6/5}$ & $-6\sqrt{6/5}$ & $2/5\sqrt{2/15}$ & $2/5\sqrt{2/15}$ & $2\sqrt{30}$ \\ \hline
$\Sigma^{*+}\pi^-$ & $-\sqrt{3/10}$ & $\sqrt{3/10}$ & $-6\sqrt{6/5}$ & $-2/5\sqrt{6/5}$ & $-2/5\sqrt{6/5}$& $2\sqrt{30}$ \\ \hline
$\Sigma^{*0}\eta$ & $\sqrt{2/5}$ & $3/\sqrt{10}$ & $0$ & $4/5\sqrt{2/5}$ & $4/5\sqrt{2/5}$ & $12\sqrt{10}$ \\ \hline
$\Sigma^{*-}\pi^+$ & $-\sqrt{3/10}$ & $\sqrt{3/10}$ & $-6\sqrt{6/5}$ & $-2/5\sqrt{6/5}$ & $-2/5\sqrt{6/5}$ & $2\sqrt{30}$ \\ \hline
$\Xi^{*0}K^0$ & $\sqrt{2/15}$ & $\sqrt{3/10}$ & $-6\sqrt{6/5}$ & $4/5\sqrt{2/15}$ & $4/5\sqrt{2/15}$ & $-4\sqrt{30}$ \\ \hline
$\Xi^{*-}K^+$ & $\sqrt{2/15}$ & $\sqrt{3/10}$ & $-6\sqrt{6/5}$ & $4/5\sqrt{2/15}$ & $4/5\sqrt{2/15}$ & $-4\sqrt{30}$ \\ \hline
\end{tabular}
\end{center}

\begin{center}
\begin{tabular}{|c|c|c|c|c|c|c|c|c|c|} \hline
$\Sigma^-_{27}$ & $\alpha^*_{27}$ & $\beta^*_{27}$ & $\gamma^*_{27}$ & $\delta^*_{27}$ & $\epsilon^*_{27}$& $\zeta^*_{27}$ \\ \hline
$\Delta^0K^-$ & $-1/2\sqrt{15}$ & $\sqrt{3/5}$ & $6\sqrt{3/5}$ & $-2/5\sqrt{15}$ & $-2/5\sqrt{15}$ & $-2\sqrt{15}$ \\ \hline
$\Delta^-\overline{K^0}$ & $-1/2\sqrt{5}$ & $3/\sqrt{5}$ & $18/\sqrt{5}$ & $-2/5\sqrt{5}$ & $-2/5\sqrt{5}$ & $-6\sqrt{5}$ \\ \hline
$\Sigma^{*0}\pi^-$ & $\sqrt{3/10}$ & $-\sqrt{3/10}$ & $6\sqrt{6/5}$ & $2/5\sqrt{6/5}$ & $2/5\sqrt{6/5}$ & $-2\sqrt{30}$ \\ \hline
$\Sigma^{*-}\pi^0$ & $-\sqrt{3/10}$ & $\sqrt{3/10}$ & $-6\sqrt{6/5}$ & $-2/5\sqrt{6/5}$ & $-2/5\sqrt{6/5}$ & $2\sqrt{30}$ \\ \hline
$\Sigma^{*-}\eta$ & $-\sqrt{2/5}$ & $-3/\sqrt{10}$ & $0$ & $-4/5\sqrt{2/5}$ & $-4/5\sqrt{2/5}$ & $-12\sqrt{10}$ \\ \hline
$\Xi^{*-}K^0$ & $-2/\sqrt{15}$ & $-\sqrt{3/5}$ & $12\sqrt{3/5}$ & $-8/5\sqrt{15}$ & $-8/5\sqrt{15}$ & $8\sqrt{15}$ \\ \hline
\end{tabular}
\end{center}

\begin{center}
\begin{tabular}{|c|c|c|c|c|c|c|} \hline
$\Lambda_{27}$ & $\alpha^*_{27}$ & $\beta^*_{27}$ & $\gamma^*_{27}$ & $\delta^*_{27}$ & $\epsilon^*_{27}$& $\zeta^*_{27}$ \\ \hline
$\Sigma^{*+}\pi^-$ & $-2/3\sqrt{2/5}$ & $9/2\sqrt{10}$ & $0$ & $-32/15\sqrt{2/5}$ & $-32/15\sqrt{2/5}$ & $0$ \\ \hline
$\Sigma^{*0}\pi^0$ & $2/3\sqrt{2/5}$ & $-9/2\sqrt{10}$ & $0$ & $32/15\sqrt{2/5}$ & $32/15\sqrt{2/5}$ & $0$ \\ \hline
$\Sigma^{*-}\pi^+$ & $2/3\sqrt{2/5}$ & $-9/2\sqrt{10}$ & $0$  & $32/15\sqrt{2/5}$ & $32/15\sqrt{2/5}$ & $0$ \\ \hline
$\Xi^{*0}K^0$ & $\sqrt{2/5}$ & $9/2\sqrt{10}$ & $0$ & $16/5\sqrt{2/5}$ & $16/5\sqrt{2/5}$ & $0$ \\ \hline
$\Xi^{*-}K^+$ & $-\sqrt{2/5}$ & $-9/2\sqrt{10}$ & $0$ & $-16/5\sqrt{2/5}$ & $-16/5\sqrt{2/5}$ & $0$  \\ \hline
\end{tabular}
\end{center}

\begin{center}
\begin{tabular}{|c|c|c|c|c|c|c|} \hline
$\Xi^{'+}_{27}$ & $\alpha^*_{27}$ & $\beta^*_{27}$ & $\gamma^*_{27}$ & $\delta^*_{27}$ & $\epsilon^*_{27}$& $\zeta^*_{27}$ \\ \hline
$\Sigma^{*+}\overline{K^0}$ & $\sqrt{2/3}$ & $0$ & $0$ & $-4\sqrt{2/3}$ & $2\sqrt{2/3}$ & $24\sqrt{6}$ \\ \hline
$\Xi^{*0}\pi^+$ & $-\sqrt{2/3}$ & $0$ & $0$ & $4\sqrt{2/3}$ & $-2\sqrt{2/3}$ & $24\sqrt{6}$ \\ \hline
\end{tabular}
\end{center}

\begin{center}
\begin{tabular}{|c|c|c|c|c|c|c|} \hline
$\Xi^{'0}_{27}$ & $\alpha^*_{27}$ & $\beta^*_{27}$ & $\gamma^*_{27}$ & $\delta^*_{27}$ & $\epsilon^*_{27}$& $\zeta^*_{27}$ \\ \hline
$\Sigma^{*+}K^-$ & $\sqrt{2}/3$ & $0$ & $0$ & $-4\sqrt{2}/3$ & $2\sqrt{2}/3$ & $24\sqrt{2}$ \\ \hline
$\Sigma^{*0}\overline{K^0}$ & $-2/3$ & $0$ & $0$ & $8/3$ & $-4/3$ & $-48$ \\ \hline
$\Xi^{*0}\pi^0$ & $-2/3$ & $0$ & $0$ & $8/3$ & $-4/3$ & $48$ \\ \hline
$\Xi^{*-}\pi^+$ & $\sqrt{2}/3$ & $0$ & $0$ & $-4\sqrt{2}/3$ & $2\sqrt{2}/3$ & $-24\sqrt{2}$ \\ \hline
\end{tabular}
\end{center}

\begin{center}
\begin{tabular}{|c|c|c|c|c|c|c|} \hline
$\Xi^{'-}_{27}$ & $\alpha^*_{27}$ & $\beta^*_{27}$ & $\gamma^*_{27}$ & $\delta^*_{27}$ & $\epsilon^*_{27}$& $\zeta^*_{27}$ \\ \hline
$\Sigma^{*0}K^-$ & $-2/3$ & $0$ & $0$ & $8/3$ & $-4/3$ & $-48$ \\ \hline
$\Sigma^{*-}\overline{K^0}$ &  $\sqrt{2}/3$ & $0$ & $0$ & $-4\sqrt{2}/3$ & $2\sqrt{2}/3$ & $24\sqrt{2}$ \\ \hline
$\Xi^{*0}\pi^-$ & $\sqrt{2}/3$ & $0$ & $0$ & $-4\sqrt{2}/3$ & $2\sqrt{2}/3$ & $-24\sqrt{2}$  \\ \hline
$\Xi^{*-}\pi^0$ & $2/3$ & $0$ & $0$ & $-8/3$ & $4/3$ & $-48$ \\ \hline
\end{tabular}
\end{center}

\begin{center}
\begin{tabular}{|c|c|c|c|c|c|c|} \hline
$\Xi^{'--}_{27}$ & $\alpha^*_{27}$ & $\beta^*_{27}$ & $\gamma^*_{27}$ & $\delta^*_{27}$ & $\epsilon^*_{27}$& $\zeta^*_{27}$ \\ \hline
$\Sigma^{*-}K^-$ & $-\sqrt{2/3}$ & $0$ & $0$ & $4\sqrt{2/3}$ & $-2\sqrt{2/3}$ & $-24\sqrt{6}$ \\ \hline
$\Xi^{*-}\pi^-$ & $\sqrt{2/3}$ & $0$ & $0$ & $-4\sqrt{2/3}$ & $2\sqrt{2/3}$ & $-24\sqrt{6}$ \\ \hline
\end{tabular}
\end{center}

\begin{center}
\begin{tabular}{|c|c|c|c|c|c|c|} \hline
$\Xi^{0}_{27}$ & $\alpha^*_{27}$ & $\beta^*_{27}$ & $\gamma^*_{27}$ & $\delta^*_{27}$ & $\epsilon^*_{27}$& $\zeta^*_{27}$ \\ \hline
$\Sigma^{*+}K^-$ & $1/3\sqrt{2/5}$ & $-3/\sqrt{10}$ & $-9\sqrt{2/5}$ & $-2/15\sqrt{2/5}$ & $28/15\sqrt{2/5}$ & $6\sqrt{10}$ \\ \hline
$\Sigma^{*0}\overline{K^0}$ & $1/3\sqrt{5}$ & $-3/2\sqrt{5}$ & $-9/\sqrt{5}$ & $-2/15\sqrt{5}$ & $28/15\sqrt{5}$ & $6\sqrt{5}$ \\ \hline
$\Xi^{*0}\pi^0$ & $-7/6\sqrt{5}$ & $3/2\sqrt{5}$ & $-9/2\sqrt{5}$ & $7/15\sqrt{5}$ & $-98/15\sqrt{5}$ & $3\sqrt{5}$ \\ \hline
$\Xi^{*0}\eta$ & $1/2\sqrt{3/5}$ & $3/2\sqrt{3/5}$ & $9/2\sqrt{3/5}$ & $-1/5\sqrt{3/5}$ & $14/5\sqrt{3/5}$ & $9\sqrt{15}$ \\ \hline
$\Xi^{*-}\pi^+$ & $-7/3\sqrt{10}$ & $3/\sqrt{10}$ & $-9/\sqrt{10}$ & $7/15\sqrt{2/5}$ & $-98/15\sqrt{2/5}$ & $3\sqrt{10}$ \\ \hline
$\Omega^-K^+$ & $\sqrt{3/10}$ & $3\sqrt{3/10}$ & $-9\sqrt{3/10}$ & $-1/5\sqrt{6/5}$ & $14/5\sqrt{6/5}$ & $-3\sqrt{30}$ \\ \hline
\end{tabular}
\end{center}

\begin{center}
\begin{tabular}{|c|c|c|c|c|c|c|} \hline
$\Xi^{-}_{27}$ & $\alpha^*_{27}$ & $\beta^*_{27}$ & $\gamma^*_{27}$ & $\delta^*_{27}$ & $\epsilon^*_{27}$& $\zeta^*_{27}$ \\ \hline
$\Sigma^{*0}K^-$ & $-1/3\sqrt{5}$ & $3/2\sqrt{5}$ & $9/\sqrt{5}$ & $2/15\sqrt{5}$ & $-28/15\sqrt{5}$ & $-6\sqrt{5}$\\ \hline
$\Sigma^{*-}\overline{K^0}$ &  $-1/3\sqrt{2/5}$ & $3/\sqrt{10}$ & $9\sqrt{2/5}$ & $2/15\sqrt{2/5}$ & $-28/15\sqrt{2/5}$ & $-6\sqrt{10}$ \\ \hline
$\Xi^{*0}\pi^-$ & $7/3\sqrt{10}$ & $-3/\sqrt{10}$ & $9/\sqrt{10}$ & $-7/15\sqrt{2/5}$ & $98/15\sqrt{2/5}$ & $-3\sqrt{10}$ \\ \hline
$\Xi^{*-}\pi^0$ & $-7/6\sqrt{5}$ & $3/2\sqrt{5}$ & $-9/2\sqrt{5}$ & $7/15\sqrt{5}$ & $-98/15\sqrt{5}$ & $3\sqrt{5}$ \\ \hline
$\Xi^{*-}\eta$ & $-1/2\sqrt{3/5}$ & $-3/2\sqrt{3/5}$ & $-9/2\sqrt{3/5}$ & $1/5\sqrt{3/5}$ & $-14/5\sqrt{3/5}$ & $-9\sqrt{15}$ \\ \hline 
$\Omega^-K^0$ & $-\sqrt{3/10}$ & $-3\sqrt{3/10}$ & $9\sqrt{3/10}$ & $1/5\sqrt{6/5}$ & $-14/5\sqrt{6/5}$ & $3\sqrt{30}$ \\ \hline
\end{tabular}
\end{center}

\begin{center}
\begin{tabular}{|c|c|c|c|c|c|c|} \hline
$\Omega^{0}_{27}$ & $\alpha^*_{27}$ & $\beta^*_{27}$ & $\gamma^*_{27}$ & $\delta^*_{27}$ & $\epsilon^*_{27}$& $\zeta^*_{27}$ \\ \hline
$\Xi^{*0}\overline{K^0}$ & $\sqrt{1/3}$ & $0$ & $0$ & $-4/\sqrt{3}$ & $8/\sqrt{3}$ & $36\sqrt{3}$ \\ \hline
$\Omega^-\pi^+$  & $-1$ & $0$ & $0$ & $4$ & $-8$ & $36$ \\ \hline
\end{tabular}
\end{center}

\begin{center}
\begin{tabular}{|c|c|c|c|c|c|c|} \hline
$\Omega^{-}_{27}$ & $\alpha^*_{27}$ & $\beta^*_{27}$ & $\gamma^*_{27}$ & $\delta^*_{27}$ & $\epsilon^*_{27}$& $\zeta^*_{27}$ \\ \hline
$\Xi^{*0}K^-$ & $\sqrt{1/6}$ & $0$ & $0$ & $-2\sqrt{2/3}$ & $4\sqrt{2/3}$ & $18\sqrt{6}$ \\ \hline
$\Xi^{*-}\overline{K^0}$ & $-\sqrt{1/6}$ & $0$ & $0$ & $2\sqrt{2/3}$ & $-4\sqrt{2/3}$ & $-18\sqrt{6}$ \\ \hline
$\Omega^-\pi^0$ & $-1$ & $0$ & $0$ & $4$ & $-8$ & $36$ \\ \hline
\end{tabular}
\end{center}

\begin{center}
\begin{tabular}{|c|c|c|c|c|c|c|} \hline
$\Omega^{--}_{27}$ & $\alpha^*_{27}$ & $\beta^*_{27}$ & $\gamma^*_{27}$ & $\delta^*_{27}$ & $\epsilon^*_{27}$& $\zeta^*_{27}$ \\ \hline
$\Xi^{*-}K^-$ & $-\sqrt{1/3}$ & $0$ & $0$ & $4/\sqrt{3}$ & $-8/\sqrt{3}$ & $-36\sqrt{3}$ \\ \hline
$\Omega^-\pi^-$  & $1$ & $0$ & $0$ & $-4$ & $8$ & $-36$ \\ \hline
\end{tabular}
\end{center}

\section{Decay Amplitudes of $35$ Pentaquark Multiplet}
\label{35table}
Below we list the decays of the thirtyfive states that form this
irreducible representation of $SU(3)$ in terms of one parameter,
$\alpha^*_{35}$, at zeroth order in $SU(3)$ symmetry breaking, plus four more
parameters, $\beta^*_{35}$, $\gamma^*_{35}$, $\delta^*_{35}$, and 
$\epsilon^*_{35}$ of order $m_s$. As was the case with the $\mathbf{27}$, 
there are several new ``exotic decays''. A truly exotic example is
$\Phi^{--}_{35}\to \Omega^-K^-$ with final state quark content $ssss\bar u$.

\begin{center}
\begin{tabular}{|c|c|c|c|c|c|c|} \hline
$\Theta^{+++}_{35}$ & $\alpha^*_{35}$ & $\beta^*_{35}$ & $\gamma^*_{35}$ & $\delta^*_{35}$ & $\epsilon^*_{35}$ \\ \hline
$\Delta^{++}K^+$ & $1$ & $0$ & $0$ & $-8$ & $16$ \\ \hline
\end{tabular}
\end{center}

\begin{center}
\begin{tabular}{|c|c|c|c|c|c|c|} \hline
$\Theta^{++}_{35}$ & $\alpha^*_{35}$ & $\beta^*_{35}$ & $\gamma^*_{35}$ & $\delta^*_{35}$ & $\epsilon^*_{35}$ \\ \hline
$\Delta^{++}K^0$ & $1/2$ & $0$ & $0$ &  $-4$ & $8$ \\ \hline
$\Delta^+K^+$ & $\sqrt{3}/2$ & $0$ & $0$ & $-4\sqrt{3}$ & $8\sqrt{3}$ \\ \hline
\end{tabular}
\end{center}

\begin{center}
\begin{tabular}{|c|c|c|c|c|c|c|} \hline
$\Theta^{+}_{35}$ & $\alpha^*_{35}$ & $\beta^*_{35}$ & $\gamma^*_{35}$ & $\delta^*_{35}$ & $\epsilon^*_{35}$ \\ \hline
$\Delta^{+}K^0$ & $\sqrt{1/2}$ & $0$ & $0$ & $-4\sqrt{2}$ & $8\sqrt{2}$ \\ \hline
$\Delta^0K^+$ & $\sqrt{1/2}$ & $0$ & $0$ &  $-4\sqrt{2}$ & $8\sqrt{2}$ \\ \hline
\end{tabular}
\end{center}

\begin{center}
\begin{tabular}{|c|c|c|c|c|c|c|} \hline
$\Theta^{0}_{35}$ & $\alpha^*_{35}$ & $\beta^*_{35}$ & $\gamma^*_{35}$ & $\delta^*_{35}$ & $\epsilon^*_{35}$ \\ \hline
$\Delta^{0}K^0$ & $\sqrt{3}/2$ & $0$ & $0$ & $-4\sqrt{3}$ & $8\sqrt{3}$ \\ \hline
$\Delta^{-}K^+$ & $1/2$ & $0$ & $0$ & $-4$ & $8$ \\ \hline
\end{tabular}
\end{center}

\begin{center}
\begin{tabular}{|c|c|c|c|c|c|c|} \hline
$\Theta^{-}_{35}$ & $\alpha^*_{35}$ & $\beta^*_{35}$ & $\gamma^*_{35}$ & $\delta^*_{35}$ & $\epsilon^*_{35}$ \\ \hline
$\Delta^{-}K^0$ & $1$ & $0$ & $0$ & $-8$ & $16$\\ \hline
\end{tabular}
\end{center}

\begin{center}
\begin{tabular}{|c|c|c|c|c|c|c|} \hline
$\Delta^{'+++}_{35}$ & $\alpha^*_{35}$ & $\beta^*_{35}$ & $\gamma^*_{35}$ & $\delta^*_{35}$ & $\epsilon^*_{35}$ \\ \hline
$\Delta^{++}\pi^+$ & $1$ & $0$ & $0$ & $4$ & $16$\\ \hline
\end{tabular}
\end{center}

\begin{center}
\begin{tabular}{|c|c|c|c|c|c|c|} \hline
$\Delta^{'++}_{35}$ & $\alpha^*_{35}$ & $\beta^*_{35}$ & $\gamma^*_{35}$ & $\delta^*_{35}$ & $\epsilon^*_{35}$ \\ \hline
$\Delta^{++}\pi^0$ & $\sqrt{2/5}$ & $0$ & $0$ & $4\sqrt{2/5}$ & $16\sqrt{2/5}$\\ \hline
$\Delta^{+}\pi^+$ & $-\sqrt{3/5}$ & $0$ & $0$ & $-4\sqrt{3/5}$ & $-16\sqrt{3/5}$\\ \hline
\end{tabular}
\end{center}

\begin{center}
\begin{tabular}{|c|c|c|c|c|c|c|} \hline
$\Delta^{'+}_{35}$ & $\alpha^*_{35}$ & $\beta^*_{35}$ & $\gamma^*_{35}$ & $\delta^*_{35}$ & $\epsilon^*_{35}$ \\ \hline
$\Delta^{++}\pi^-$ & $\sqrt{1/10}$ & $0$ & $0$ & $2\sqrt{2/5}$ & $8\sqrt{2/5}$\\ \hline
$\Delta^{+}\pi^0$ & $\sqrt{3/5}$ & $0$ & $0$ & $4\sqrt{3/5}$ & $16\sqrt{3/5}$\\ \hline
$\Delta^{0}\pi^+$ & $-\sqrt{3/10}$ & $0$ & $0$ & $-2\sqrt{6/5}$ & $-8\sqrt{6/5}$\\ \hline
\end{tabular}
\end{center}

\begin{center}
\begin{tabular}{|c|c|c|c|c|c|c|} \hline
$\Delta^{'0}_{35}$ & $\alpha^*_{35}$ & $\beta^*_{35}$ & $\gamma^*_{35}$ & $\delta^*_{35}$ & $\epsilon^*_{35}$ \\ \hline
$\Delta^{+}\pi^-$ & $\sqrt{3/10}$ & $0$ & $0$ & $2\sqrt{6/5}$ & $8\sqrt{6/5}$\\ \hline
$\Delta^{0}\pi^0$ & $\sqrt{3/5}$ & $0$ & $0$ & $4\sqrt{3/5}$ & $16\sqrt{3/5}$\\ \hline
$\Delta^{-}\pi^+$ & $-\sqrt{1/10}$ & $0$ & $0$ & $-2\sqrt{2/5}$ & $-8\sqrt{2/5}$\\ \hline
\end{tabular}
\end{center}

\begin{center}
\begin{tabular}{|c|c|c|c|c|c|c|} \hline
$\Delta^{'-}_{35}$ & $\alpha^*_{35}$ & $\beta^*_{35}$ & $\gamma^*_{35}$ & $\delta^*_{35}$ & $\epsilon^*_{35}$ \\ \hline
$\Delta^{0}\pi^-$ & $\sqrt{3/5}$ & $0$ & $0$ & $4\sqrt{3/5}$ & $16\sqrt{3/5}$\\ \hline
$\Delta^{-}\pi^0$ & $\sqrt{2/5}$ & $0$ & $0$ & $4\sqrt{2/5}$ & $16\sqrt{2/5}$\\ \hline
\end{tabular}
\end{center}

\begin{center}
\begin{tabular}{|c|c|c|c|c|c|c|} \hline
$\Delta^{'--}_{35}$ & $\alpha^*_{35}$ & $\beta^*_{35}$ & $\gamma^*_{35}$ & $\delta^*_{35}$ & $\epsilon^*_{35}$ \\ \hline
$\Delta^{-}\pi^-$ & $1$ & $0$ & $0$ & $4$ & $16$\\ \hline
\end{tabular}
\end{center}

\begin{center}
\begin{tabular}{|c|c|c|c|c|c|c|} \hline
$\Delta^{++}_{35}$ & $\alpha^*_{35}$ & $\beta^*_{35}$ & $\gamma^*_{35}$ & $\delta^*_{35}$ & $\epsilon^*_{35}$ \\ \hline
$\Delta^{++}\pi^0$ & $1/4\sqrt{3/5}$ & $3\sqrt{15}/4$ & $-\sqrt{15}/2$ & $-3/2\sqrt{3/5}$ & $3/2\sqrt{3/5}$\\ \hline
$\Delta^{++}\eta$ & $\sqrt{5}/4$ & $3\sqrt{5}/4$ & $3\sqrt{5}/2$ & $-3\sqrt{5}/2$ & $3\sqrt{5}/2$ \\ \hline
$\Delta^+\pi^+$ & $1/2\sqrt{10}$ & $3/2\sqrt{5/2}$ & $-\sqrt{5/2}$ & $-3/\sqrt{10}$ & $3/\sqrt{10}$ \\ \hline
$\Sigma^{*+}K^+$ & $-1/2\sqrt{5/2}$ & $3/2\sqrt{5/2}$ & $\sqrt{5/2}$ & $3\sqrt{5/2}$ & $-3\sqrt{5/2}$ \\ \hline
\end{tabular}
\end{center}

\begin{center}
\begin{tabular}{|c|c|c|c|c|c|c|} \hline
$\Delta^{+}_{35}$ & $\alpha^*_{35}$ & $\beta^*_{35}$ & $\gamma^*_{35}$ & $\delta^*_{35}$ & $\epsilon^*_{35}$ \\ \hline
$\Delta^{++}\pi^-$ & $1/2\sqrt{10}$ & $3/2\sqrt{5/2}$ & $-\sqrt{5/2}$ & $-3/\sqrt{10}$ & $3/\sqrt{10}$ \\ \hline
$\Delta^{+}\pi^0$ & $1/4\sqrt{15}$ & $\sqrt{15}/4$ & $-1/2\sqrt{5/3}$ & $-1/2\sqrt{3/5}$ & $1/2\sqrt{3/5}$ \\ \hline
$\Delta^+\eta$ & $\sqrt{5}/4$ & $3\sqrt{5}/4$ & $3\sqrt{5}/2$ & $-3\sqrt{5}/2$ & $3\sqrt{5}/2$ \\ \hline
$\Delta^0\pi^+$ & $\sqrt{1/30}$ & $\sqrt{15/2}$ & $-\sqrt{10/3}$ & $-\sqrt{6/5}$ & $\sqrt{6/5}$ \\ \hline
$\Sigma^{*+}K^0$ & $-1/2\sqrt{5/6}$ & $1/2\sqrt{15/2}$ & $\sqrt{5/6}$ & $\sqrt{15/2}$ & $-\sqrt{15/2}$ \\ \hline
$\Sigma^{*0}K^+$ & $-1/2\sqrt{5/3}$ & $\sqrt{15}/2$ & $\sqrt{5/3}$ & $\sqrt{15}$ & $-\sqrt{15}$ \\ \hline
\end{tabular}
\end{center}

\begin{center}
\begin{tabular}{|c|c|c|c|c|c|c|} \hline
$\Delta^{0}_{35}$ & $\alpha^*_{35}$ & $\beta^*_{35}$ & $\gamma^*_{35}$ & $\delta^*_{35}$ & $\epsilon^*_{35}$ \\ \hline
$\Delta^{+}\pi^-$ & $\sqrt{1/30}$ & $\sqrt{15/2}$  & $-\sqrt{10/3}$ & $-\sqrt{6/5}$ & $\sqrt{6/5}$ \\ \hline
$\Delta^{0}\pi^0$ & $-1/4\sqrt{15}$ & $-\sqrt{15}/4$ & $1/2\sqrt{5/3}$ & $1/2\sqrt{3/5}$ & $-1/2\sqrt{3/5}$ \\ \hline
$\Delta^0\eta$ & $\sqrt{5}/4$ & $3\sqrt{5}/4$ & $3\sqrt{5}/2$ & $-3\sqrt{5}/2$ & $3\sqrt{5}/2$ \\ \hline
$\Delta^-\pi^+$ & $1/2\sqrt{10}$ & $3/2\sqrt{5/2}$ & $-\sqrt{5/2}$ & $-3/\sqrt{10}$ & $3/\sqrt{10}$ \\ \hline
$\Sigma^{*0}K^0$ & $-1/2\sqrt{5/3}$ & $\sqrt{15}/2$ & $\sqrt{5/3}$ & $\sqrt{15}$ & $-\sqrt{15}$ \\ \hline
$\Sigma^{*-}K^+$ & $-1/2\sqrt{5/6}$ & $1/2\sqrt{15/2}$ & $\sqrt{5/6}$ & $\sqrt{15/2}$ & $-\sqrt{15/2}$ \\ \hline
\end{tabular}
\end{center}

\begin{center}
\begin{tabular}{|c|c|c|c|c|c|c|} \hline
$\Delta^{-}_{35}$ & $\alpha^*_{35}$ & $\beta^*_{35}$ & $\gamma^*_{35}$ & $\delta^*_{35}$ & $\epsilon^*_{35}$ \\ \hline
$\Delta^{0}\pi^-$ & $1/2\sqrt{10}$ & $3/2\sqrt{5/2}$ & $-\sqrt{5/2}$ & $-3/\sqrt{10}$ & $3/\sqrt{10}$ \\ \hline
$\Delta^-\pi^0$ & $-1/4\sqrt{3/5}$ & $-3\sqrt{15}/4$ & $\sqrt{15}/2$ & $3/2\sqrt{3/5}$ & $-3/2\sqrt{3/5}$ \\ \hline
$\Delta^-\eta$ & $\sqrt{5}/4$ & $3\sqrt{5}/4$ & $3\sqrt{5}/2$ & $-3\sqrt{5}/2$ & $3\sqrt{5}/2$ \\ \hline
$\Sigma^{*-}K^0$ & $-1/2\sqrt{5/2}$ & $3/2\sqrt{5/2}$ & $\sqrt{5/2}$ & $3\sqrt{5/2}$ & $-3\sqrt{5/2}$ \\ \hline
\end{tabular}
\end{center}

\begin{center}
\begin{tabular}{|c|c|c|c|c|c|c|} \hline
$\Sigma^{'++}_{35}$ & $\alpha^*_{35}$ & $\beta^*_{35}$ & $\gamma^*_{35}$ & $\delta^*_{35}$ & $\epsilon^*_{35}$ \\ \hline
$\Delta^{++}\overline{K^0}$ & $1/2$ & $0$ & $6$ & $2$ & $2$ \\ \hline
$\Sigma^{*+}\pi^+$ & $\sqrt{3}/2$ & $0$ & $-2\sqrt{3}$ & $2\sqrt{3}$ & $2\sqrt{3}$ \\ \hline
\end{tabular}
\end{center}

\begin{center}
\begin{tabular}{|c|c|c|c|c|c|c|} \hline
$\Sigma^{'+}_{35}$ & $\alpha^*_{35}$ & $\beta^*_{35}$ & $\gamma^*_{35}$ & $\delta^*_{35}$ & $\epsilon^*_{35}$ \\ \hline
$\Delta^{++}K^-$ & $1/4$ & $0$ & $3$ & $1$ & $1$ \\ \hline
$\Delta^{+}\overline{K^0}$ & $-\sqrt{3}/4$ & $0$ & $-3\sqrt{3}$ & $-\sqrt{3}$ & $-\sqrt{3}$ \\ \hline
$\Sigma^{*+}\pi^0$ & $1/2\sqrt{3/2}$ & $0$ & $-\sqrt{6}$ & $\sqrt{6}$ & $\sqrt{6}$ \\ \hline
$\Sigma^{*0}\pi^+$ & $-1/2\sqrt{3/2}$ & $0$ & $\sqrt{6}$ & $-\sqrt{6}$ & $-\sqrt{6}$ \\ \hline
\end{tabular}
\end{center}

\begin{center}
\begin{tabular}{|c|c|c|c|c|c|c|} \hline
$\Sigma^{'0}_{35}$ & $\alpha^*_{35}$ & $\beta^*_{35}$ & $\gamma^*_{35}$ & $\delta^*_{35}$ & $\epsilon^*_{35}$ \\ \hline
$\Delta^{+}K^-$ & $1/2\sqrt{2}$ & $0$ & $3\sqrt{2}$ & $\sqrt{2}$ & $\sqrt{2}$ \\ \hline
$\Delta^{0}\overline{K^0}$ & $-1/2\sqrt{2}$ & $0$ & $-3\sqrt{2}$ & $-\sqrt{2}$ & $-\sqrt{2}$ \\ \hline
$\Sigma^{*+}\pi^-$ & $1/2\sqrt{2}$ & $0$ & $-\sqrt{2}$ & $\sqrt{2}$ & $\sqrt{2}$ \\ \hline
$\Sigma^{*0}\pi^0$ & $\sqrt{1/2}$ & $0$ & $-2\sqrt{2}$ & $2\sqrt{2}$ & $2\sqrt{2}$ \\ \hline
$\Sigma^{*-}\pi^+$ & $-1/2\sqrt{2}$ & $0$ & $\sqrt{2}$ & $-\sqrt{2}$ & $-\sqrt{2}$ \\ \hline
\end{tabular}
\end{center}

\begin{center}
\begin{tabular}{|c|c|c|c|c|c|c|} \hline
$\Sigma^{'-}_{35}$ & $\alpha^*_{35}$ & $\beta^*_{35}$ & $\gamma^*_{35}$ & $\delta^*_{35}$ & $\epsilon^*_{35}$ \\ \hline
$\Delta^{0}K^-$ & $\sqrt{3}/4$ & $0$ & $3\sqrt{3}$ & $\sqrt{3}$ & $\sqrt{3}$ \\ \hline
$\Delta^{-}\overline{K^0}$ & $-1/4$ & $0$ & $-3$ & $-1$ & $-1$ \\ \hline
$\Sigma^{*0}\pi^-$ & $1/2\sqrt{3/2}$ & $0$ & $-\sqrt{6}$ & $\sqrt{6}$ & $\sqrt{6}$ \\ \hline
$\Sigma^{*-}\pi^0$ & $1/2\sqrt{3/2}$ & $0$ & $-\sqrt{6}$ & $\sqrt{6}$ & $\sqrt{6}$ \\ \hline
\end{tabular}
\end{center}

\begin{center}
\begin{tabular}{|c|c|c|c|c|c|c|} \hline
$\Sigma^{'--}_{35}$ & $\alpha^*_{35}$ & $\beta^*_{35}$ & $\gamma^*_{35}$ & $\delta^*_{35}$ & $\epsilon^*_{35}$ \\ \hline
$\Delta^{-}K^-$ & $1/2$ & $0$ & $6$ & $2$ & $2$  \\ \hline
$\Sigma^{*-}\pi^-$  & $\sqrt{3}/2$ & $0$ &  $-2\sqrt{3}$& $2\sqrt{3}$ & $2\sqrt{3}$\\ \hline
\end{tabular}
\end{center}

\begin{center}
\begin{tabular}{|c|c|c|c|c|c|c|} \hline
$\Sigma^{+}_{35}$ & $\alpha^*_{35}$ & $\beta^*_{35}$ & $\gamma^*_{35}$ & $\delta^*_{35}$ & $\epsilon^*_{35}$ \\ \hline
$\Delta^{++}K^-$ & $1/4$ & $3$ & $1$ & $-1$ & $-1$  \\ \hline
$\Delta^{+}\overline{K^0}$ & $1/4\sqrt{3}$ & $\sqrt{3}$ & $\sqrt{1/3}$ & $-\sqrt{1/3}$  & $-\sqrt{1/3}$ \\ \hline
$\Sigma^{*+}\pi^0$  & $1/2\sqrt{6}$ & $\sqrt{6}$ &  $-\sqrt{6}$& $-\sqrt{2/3}$ & $-\sqrt{2/3}$\\ \hline
$\Sigma^{*+}\eta$  & $\sqrt{1/2}$ & $0$ & $2\sqrt{2}$ &  $-2\sqrt{2}$& $-2\sqrt{2}$ \\ \hline
$\Sigma^{*0}\pi^+$ & $1/2\sqrt{6}$ & $\sqrt{6}$ & $-\sqrt{6}$ & $-\sqrt{2/3}$ & $-\sqrt{2/3}$ \\ \hline
$\Xi^{*0}K^+$ & $-\sqrt{1/3}$ & $2\sqrt{3}$ & $4/\sqrt{3}$ & $4/\sqrt{3}$ & $4/\sqrt{3}$ \\ \hline
\end{tabular}
\end{center}

\begin{center}
\begin{tabular}{|c|c|c|c|c|c|c|} \hline
$\Sigma^{0}_{35}$ & $\alpha^*_{35}$ & $\beta^*_{35}$ & $\gamma^*_{35}$ & $\delta^*_{35}$ & $\epsilon^*_{35}$ \\ \hline
$\Delta^+K^-$ & $1/2\sqrt{6}$ & $\sqrt{6}$ & $\sqrt{2/3}$ & $-\sqrt{2/3}$ & $-\sqrt{2/3}$ \\ \hline
$\Delta^{0}\overline{K^0}$ & $1/2\sqrt{6}$ & $\sqrt{6}$ & $\sqrt{2/3}$ & $-\sqrt{2/3}$  & $-\sqrt{2/3}$ \\ \hline
$\Sigma^{*+}\pi^-$  & $1/2\sqrt{6}$ & $\sqrt{6}$ &  $-\sqrt{6}$& $-\sqrt{2/3}$ & $-\sqrt{2/3}$\\ \hline
$\Sigma^{*0}\eta$  & $\sqrt{1/2}$ & $0$ & $2\sqrt{2}$ &  $-2\sqrt{2}$& $-2\sqrt{2}$ \\ \hline
$\Sigma^{*-}\pi^+$ & $1/2\sqrt{6}$ & $\sqrt{6}$ & $-\sqrt{6}$ & $-\sqrt{2/3}$ & $-\sqrt{2/3}$ \\ \hline
$\Xi^{*0}K^0$ & $-\sqrt{1/6}$ & $\sqrt{6}$ & $2\sqrt{2/3}$ & $2\sqrt{2/3}$ & $2\sqrt{2/3}$ \\ \hline
$\Xi^{*-}K^+$ & $-\sqrt{1/6}$ & $\sqrt{6}$ & $2\sqrt{2/3}$ & $2\sqrt{2/3}$ & $2\sqrt{2/3}$ \\ \hline
\end{tabular}
\end{center}

\begin{center}
\begin{tabular}{|c|c|c|c|c|c|c|} \hline
$\Sigma^{-}_{35}$ & $\alpha^*_{35}$ & $\beta^*_{35}$ & $\gamma^*_{35}$ & $\delta^*_{35}$ & $\epsilon^*_{35}$ \\ \hline
$\Delta^0K^-$ & $1/4\sqrt{3}$ & $\sqrt{3}$ & $\sqrt{1/3}$ & $-\sqrt{1/3}$ & $-\sqrt{1/3}$ \\ \hline
$\Delta^{-}\overline{K^0}$ & $1/4$ & $3$ & $1$ & $-1$  & $-1$ \\ \hline
$\Sigma^{*0}\pi^-$  & $1/2\sqrt{6}$ & $\sqrt{6}$ &  $-\sqrt{6}$& $-\sqrt{2/3}$ & $-\sqrt{2/3}$\\ \hline
$\Sigma^{*-}\pi^0$ & $-1/2\sqrt{6}$ & $-\sqrt{6}$ & $\sqrt{6}$ & $\sqrt{2/3}$ & $\sqrt{2/3}$ \\ \hline
$\Sigma^{*-}\eta$  & $\sqrt{1/2}$ & $0$ & $2\sqrt{2}$ &  $-2\sqrt{2}$& $-2\sqrt{2}$ \\ \hline
$\Xi^{*-}K^0$ & $-\sqrt{1/3}$ & $2\sqrt{3}$ & $4/\sqrt{3}$ & $4/\sqrt{3}$ & $4/\sqrt{3}$ \\ \hline
\end{tabular}
\end{center}

\begin{center}
\begin{tabular}{|c|c|c|c|c|c|c|} \hline
$\Xi^{'+}_{35}$ & $\alpha^*_{35}$ & $\beta^*_{35}$ & $\gamma^*_{35}$ & $\delta^*_{35}$ & $\epsilon^*_{35}$ \\ \hline
$\Sigma^{*+}\overline{K^0}$  &$\sqrt{1/2}$ &   $0$ &  $4\sqrt{2}$& $2\sqrt{2}$ & $-4\sqrt{2}$\\ \hline
$\Xi^{*0}\pi^+$ & $\sqrt{1/2}$ & $0$ & $-4\sqrt{2}$ & $2\sqrt{2}$ & $-4\sqrt{2}$ \\ \hline
\end{tabular}
\end{center}

\begin{center}
\begin{tabular}{|c|c|c|c|c|c|c|} \hline
$\Xi^{'0}_{35}$ & $\alpha^*_{35}$ & $\beta^*_{35}$ & $\gamma^*_{35}$ & $\delta^*_{35}$ & $\epsilon^*_{35}$ \\ \hline
$\Sigma^{*+}K^-$  &$\sqrt{1/6}$ &  $0$ &  $4\sqrt{2/3}$& $2\sqrt{2/3}$ & $-4\sqrt{2/3}$\\ \hline
$\Sigma^{*0}\overline{K^0}$ & $-\sqrt{1/3}$ &  $0$ &  $-8/\sqrt{3}$& $-4/\sqrt{3}$ & $8/\sqrt{3}$\\ \hline
$\Xi^{*0}\pi^0$ & $\sqrt{1/3}$ & $0$ & $-8/\sqrt{3}$ & $4/\sqrt{3}$ & $-8/\sqrt{3}$ \\ \hline
$\Xi^{*-}\pi^+$ & $-\sqrt{1/6}$ & $0$ & $4\sqrt{2/3}$ & $-2/\sqrt{2/3}$ & $4\sqrt{2/3}$ \\ \hline
\end{tabular}
\end{center}

\begin{center}
\begin{tabular}{|c|c|c|c|c|c|c|} \hline
$\Xi^{'-}_{35}$ & $\alpha^*_{35}$ & $\beta^*_{35}$ & $\gamma^*_{35}$ & $\delta^*_{35}$ & $\epsilon^*_{35}$ \\ \hline
$\Sigma^{*0}K^-$ & $\sqrt{1/3}$ &  $0$ &  $8/\sqrt{3}$& $4/\sqrt{3}$ & $-8/\sqrt{3}$\\ \hline
$\Sigma^{*-}\overline{K^0}$ & $-\sqrt{1/6}$ &  $0$ &  $-4/\sqrt{2/3}$& $-2/\sqrt{2/3}$ & $4\sqrt{2/3}$\\ \hline
$\Xi^{*0}\pi^-$ & $\sqrt{1/6}$ & $0$ & $-4\sqrt{2/3}$ & $2\sqrt{2/3}$ & $-4\sqrt{2/3}$ \\ \hline
$\Xi^{*-}\pi^0$ & $\sqrt{1/3}$ & $0$ & $-8/\sqrt{3}$ & $4/\sqrt{3}$ & $-8/\sqrt{3}$ \\ \hline
\end{tabular}
\end{center}

\begin{center}
\begin{tabular}{|c|c|c|c|c|c|c|} \hline
$\Xi^{'--}_{35}$ & $\alpha^*_{35}$ & $\beta^*_{35}$ & $\gamma^*_{35}$ & $\delta^*_{35}$ & $\epsilon^*_{35}$ \\ \hline
$\Sigma^{*-}K^-$ & $\sqrt{1/2}$ &  $0$ &  $4\sqrt{2}$& $2\sqrt{2}$ & $-4\sqrt{2}$\\ \hline
$\Xi^{*-}\pi^-$ &  $\sqrt{1/2}$ &  $0$ &  $-4\sqrt{2}$& $2\sqrt{2}$ & $-4\sqrt{2}$\\ \hline
\end{tabular}
\end{center}

\begin{center}
\begin{tabular}{|c|c|c|c|c|c|c|} \hline
$\Xi^{0}_{35}$ & $\alpha^*_{35}$ & $\beta^*_{35}$ & $\gamma^*_{35}$ & $\delta^*_{35}$ & $\epsilon^*_{35}$ \\ \hline
$\Sigma^{*+}K^-$ & $\sqrt{1/6}$ & $3\sqrt{3/2}$& $\sqrt{2/3}$ & $-\sqrt{2/3}$ & $-7\sqrt{2/3}$\\ \hline
$\Sigma^{*0}\overline{K^0}$ & $1/2\sqrt{3}$ & $3\sqrt{3}/2$& $\sqrt{1/3}$ & $-\sqrt{1/3}$ & $-7/\sqrt{3}$\\ \hline
$\Xi^{*0}\pi^0$ & $1/4\sqrt{3}$ & $3\sqrt{3}/4$ & $-7/2\sqrt{3}$ & $-1/2\sqrt{3}$ & $-7/2\sqrt{3}$ \\ \hline
$\Xi^{*0}\eta$ & $3/4$ & $-9/4$ & $3/2$ & $-3/2$ & $-21/2$ \\ \hline
$\Xi^{*-}\pi^+$ & $1/2\sqrt{6}$ & $3/2\sqrt{3/2}$ & $-7/\sqrt{6}$ & $-\sqrt{1/6}$ & $-7/\sqrt{6}$ \\ \hline
$\Omega^-K^+$ & $-1/2\sqrt{2}$ & $9/2\sqrt{2}$ & $3/\sqrt{2}$ & $\sqrt{1/2}$ & $7/\sqrt{2}$ \\ \hline
\end{tabular}
\end{center}

\begin{center}
\begin{tabular}{|c|c|c|c|c|c|c|} \hline
$\Xi^{-}_{35}$ & $\alpha^*_{35}$ & $\beta^*_{35}$ & $\gamma^*_{35}$ & $\delta^*_{35}$ & $\epsilon^*_{35}$ \\ \hline
$\Sigma^{*0}K^-$ & $1/2\sqrt{3}$ &  $3\sqrt{3}/2$ &  $\sqrt{1/3}$& $-\sqrt{1/3}$ & $-7/\sqrt{3}$\\ \hline
$\Sigma^{*-}\overline{K^0}$ & $\sqrt{1/6}$ & $3\sqrt{3/2}$ & $\sqrt{2/3}$ & $-\sqrt{2/3}$ & $-7\sqrt{2/3}$ \\ \hline
$\Xi^{*0}\pi^-$ & $1/2\sqrt{6}$ & $3/2\sqrt{3/2}$ & $-7/\sqrt{6}$ & $-\sqrt{1/6}$ & $-7/\sqrt{6}$ \\ \hline 
$\Xi^{*-}\pi^0$ & $-1/4\sqrt{3}$ & $-3\sqrt{3}/4$ & $7/2\sqrt{3}$ & $1/2\sqrt{3}$ & $7/2\sqrt{3}$ \\ \hline
$\Xi^{*-}\eta$ & $3/4$ & $-9/4$ & $3/2$ & $-3/2$ & $-21/2$ \\ \hline
$\Omega^-K^0$ & $-1/2\sqrt{2}$ & $9/2\sqrt{2}$ & $3/\sqrt{2}$ & $\sqrt{1/2}$ & $7/\sqrt{2}$ \\ \hline
\end{tabular}
\end{center}

\begin{center}
\begin{tabular}{|c|c|c|c|c|c|c|} \hline
$\Omega^{'0}_{35}$ & $\alpha^*_{35}$ & $\beta^*_{35}$ & $\gamma^*_{35}$ & $\delta^*_{35}$ & $\epsilon^*_{35}$ \\ \hline
$\Xi^{*0}\overline{K^0}$ & $\sqrt{3}/2$ & $0$ & $2\sqrt{3}$ & $2\sqrt{3}$ & $-10\sqrt{3}$ \\ \hline
$\Omega^-\pi^+$ & $1/2$ & $0$ & $-6$ & $2$ & $-10$ \\ \hline
\end{tabular}
\end{center}

\begin{center}
\begin{tabular}{|c|c|c|c|c|c|c|} \hline
$\Omega^{'-}_{35}$ & $\alpha^*_{35}$ & $\beta^*_{35}$ & $\gamma^*_{35}$ & $\delta^*_{35}$ & $\epsilon^*_{35}$ \\ \hline
$\Xi^{*0}K^-$ & $1/2\sqrt{3/2}$ & $0$ & $\sqrt{6}$ & $\sqrt{6}$ & $-5\sqrt{6}$ \\ \hline
$\Xi^{*-}\overline{K^0}$ & $-1/2\sqrt{3/2}$ & $0$ & $-\sqrt{6}$ & $-\sqrt{6}$ & $5\sqrt{6}$ \\ \hline
$\Omega^-\pi^0$ & $1/2$ & $0$ & $-6$ & $2$ & $-10$ \\ \hline
\end{tabular}
\end{center}

\begin{center}
\begin{tabular}{|c|c|c|c|c|c|c|} \hline
$\Omega^{'--}_{35}$ & $\alpha^*_{35}$ & $\beta^*_{35}$ & $\gamma^*_{35}$ & $\delta^*_{35}$ & $\epsilon^*_{35}$ \\ \hline
$\Xi^{*-}K^-$ & $\sqrt{3}/2$ & $0$ & $2\sqrt{3}$ & $2\sqrt{3}$ & $-10\sqrt{3}$ \\ \hline
$\Omega^-\pi^-$ & $1/2$ & $0$ & $-6$ & $2$ & $-10$ \\ \hline
\end{tabular}
\end{center}

\begin{center}
\begin{tabular}{|c|c|c|c|c|c|c|} \hline
$\Omega^{-}_{35}$ & $\alpha^*_{35}$ & $\beta^*_{35}$ & $\gamma^*_{35}$ & $\delta^*_{35}$ & $\epsilon^*_{35}$ \\ \hline
$\Xi^{*0}K^-$ & $1/2$ & $3$ & $0$ & $0$ & $-12$  \\ \hline
$\Xi^{*-}\overline{K^0}$ & $1/2$ & $3$ & $0$ & $0$ & $-12$ \\ \hline
$\Omega^{-}\eta$ & $\sqrt{1/2}$ & $-3\sqrt{2}$ & $0$ & $0$ & $-12\sqrt{2}$ \\ \hline
\end{tabular}
\end{center}

\begin{center}
\begin{tabular}{|c|c|c|c|c|c|c|} \hline
$\Phi^{-}_{35}$ & $\alpha^*_{35}$ & $\beta^*_{35}$ & $\gamma^*_{35}$ & $\delta^*_{35}$ & $\epsilon^*_{35}$ \\ \hline
$\Omega^{-}\overline{K^0}$ & $1$ & $0$ & $0$ & $4$ &  $-32$ \\ \hline
\end{tabular}
\end{center}

\begin{center}
\begin{tabular}{|c|c|c|c|c|c|c|} \hline
$\Phi^{--}_{35}$ & $\alpha^*_{35}$ & $\beta^*_{35}$ & $\gamma^*_{35}$ & $\delta^*_{35}$ & $\epsilon^*_{35}$ \\ \hline
$\Omega^{-}K^-$ & $1$ & $0$ & $0$ & $4$ &  $-32$ \\ \hline
\end{tabular}
\end{center}

\section{Acknowledgments}
This work was supported in part by the Department of Energy under
Grant DE-FG03-97ER40546.

\end{document}